\documentclass[preprint,10pt]{elsarticle}
\usepackage{eurosym}
\usepackage{amsmath,amssymb}
\usepackage{graphicx}
\usepackage{hyperref}
\usepackage[utf8]{inputenc}
\usepackage{graphicx}
\usepackage{subcaption}
\usepackage{color}

\usepackage[
  a4paper,
  left=1.0cm,
  right=1.0cm,
  top=2.0cm,
  bottom=1.5cm,
  footskip=0.7cm
]{geometry}

\journal{Physica A: Statistical Mechanics and its Applications}

\begin{document}

\begin{frontmatter}


\title{Does the Power-Law Advantage in the Tails Outweigh the Global q-Gaussian Description?}

\author[inst1]{Eduardo Boor}
\ead{eduardo.boor@ufrgs.br}

\author[inst1]{Roberto da Silva\corref{cor1} }
\ead{rdasilva@if.ufrgs.br}

\author[inst1]{João Carlos Schmitt de Siqueira}
\ead{schmitt.siqueira@ufrgs.br}

\author[inst1]{J. Roberto Iglesias}
\ead{iglesias@if.ufrgs.br}

\cortext[cor1]{Corresponding author}

\address[inst1]{Institute of Physics, Federal University of Rio Grande do Sul, Porto Alegre, Rio Grande do Sul, 91509-900, Brazil}

\begin{abstract}
Financial markets are complex systems whose return distributions exhibit heavy tails, traditionally described by 
power laws. An alternative framework based on Tsallis nonextensive statistical mechanics allows both the central 
region and the tails of the distribution to be represented by a single q-Gaussian functional form. In this work, we 
systematically compare these two descriptions across different time scales, considering stock markets from Brazil 
and the United States, as well as traditional assets and cryptocurrencies. The q-Gaussian parameters are estimated 
using the method of moments ratio, whereas the power-law parameters are obtained by maximum likelihood estimation 
in the asymptotic region. The quality of the fits is assessed using goodness-of-fit measures, bootstrap resampling, 
and the Vuong test. When the analysis is restricted exclusively to the extreme tails, the power law generally 
provides a better description, as expected from the asymptotic power-law behavior of the q-Gaussian itself. 
However, this superiority is not sufficiently pronounced to outweigh the main advantage of the q-Gaussian: its 
ability to provide a consistent description of the entire distribution, including the central peak, bulk, and tails, 
over all investigated time scales. Moreover, the evolution of the parameter $q$ offers a simple and direct 
characterization of aggregational Gaussianity and of the transition between heavy- and short-tailed statistical 
regimes, without requiring separate fits for different regions of the distribution. These results indicate that, 
although the power law remains particularly suitable for describing extreme events, the q-Gaussian provides 
a broader and more practical framework for characterizing the statistical evolution of financial returns.

\end{abstract}


\end{frontmatter}


\section{Introduction}

\label{sec:intro}

The analysis of complex time series, ranging from fluid turbulence to the
high-frequency dynamics of financial markets, has repeatedly revealed
statistical properties that cannot be adequately described within the
framework of conventional Gaussian statistics. In econophysics, financial
returns are known to display a number of robust empirical regularities,
commonly referred to as stylized facts, among which heavy-tailed
distributions and aggregational Gaussianity are particularly prominent \cite%
{mandelbrot1963,fama1965,mantegna1995,cont2001,dasilva2010,ramos2000}. At short aggregation intervals, $\Delta t$, return
distributions exhibit pronounced deviations from Gaussian behavior, with a
substantially enhanced probability of extreme fluctuations. As $\Delta t$
increases, however, the distributions progressively approach a Gaussian
form. This scale-dependent evolution constitutes one of the most
characteristic statistical signatures of financial time series.

Heavy tails are traditionally associated with power-law behavior and have
consequently been extensively analyzed within this framework. Nevertheless,
identifying a genuine power law in empirical data is a subtle statistical
problem, since power-law behavior is expected only beyond a sufficiently
large threshold and its detection may depend strongly on the definition of
the asymptotic region, the estimation procedure, and the finite amount of
data available \cite{lux1996,malevergne2005,clauset2009,todorova2011}. Moreover, a pure power law
is intrinsically a description of the tail and therefore does not provide a
unified representation of the central region and the extreme fluctuations of
the return distribution.

An alternative description arises naturally from the nonextensive
statistical-mechanics formalism introduced by Tsallis \cite{tsallis1988,tsallis1999}.
Within this framework, the Gaussian distribution is generalized to the $q$%
-Gaussian, characterized by the entropic index $q$. For $q>1$, the $q$%
-Gaussian exhibits heavy tails, whereas the standard Gaussian distribution
is continuously recovered in the limit $q\rightarrow 1$. Importantly, the $q$%
-Gaussian provides a single functional form capable of describing the
central peak, the intermediate region, and the tails of the distribution. In
this sense, it offers a particularly appealing framework for financial
returns, for which both the bulk of ordinary fluctuations and the occurrence
of extreme events are relevant \cite{tsallis2003economics,queiros2005,borland2002}.

The connection between these two descriptions is especially important. For
sufficiently large fluctuations, the $q$-Gaussian itself asymptotically
approaches a power law. Therefore, if the statistical analysis is restricted
exclusively to the far-tail region, one may naturally expect a pure
power-law fit, whose parameters are optimized specifically for that region,
to perform at least as well as, and possibly better than, a $q$-Gaussian
whose parameters characterize the distribution more globally. Consequently,
the relevant question is not simply whether a power law can provide a better
fit to the most extreme observations, but whether such an improvement is
sufficiently significant to justify abandoning a distribution that
simultaneously describes the central peak, the bulk, and the tails.

This distinction becomes particularly relevant when different temporal
aggregation scales are considered. Aggregational Gaussianity implies a
progressive transformation of the return distribution from a heavy-tailed
form at short time scales toward Gaussian statistics at longer scales.
Within the nonextensive framework, this evolution can be monitored directly
through the parameter $q$: strongly non-Gaussian distributions are
associated with $q>1$, while the relaxation toward $q=1$ provides a simple
quantitative representation of the approach to the Gaussian regime \cite%
{ramos2000,tsallis1999,umarov2008}. Thus, rather than fitting different functional
forms independently to different portions of the distribution or at
different aggregation scales, the evolution of a single parameter can be
used to characterize the transition between heavy- and short-tailed
statistical regimes.

This perspective suggests that the comparison between the $q$-Gaussian and
the pure power law should be performed at two complementary levels. On the
one hand, the extreme-tail region must be analyzed separately in order to
determine whether a genuine power-law description provides a statistically
superior representation of rare events. On the other hand, the complete
distribution must also be considered, since financial dynamics cannot be
characterized exclusively by its most extreme fluctuations. A statistically
meaningful comparison must therefore distinguish between local superiority
in the asymptotic tail and the ability to provide a robust description of
the distribution as a whole across different time scales.

In this work, we address precisely this issue by systematically comparing
the $q$-Gaussian and pure power-law descriptions of financial returns over a
broad range of temporal aggregation scales. Our analysis encompasses
conventional stock markets from both Brazil and the United States, as well
as cryptocurrencies, allowing us to test the robustness of these statistical
descriptions across different markets and classes of financial assets. We
examine not only the quality of the global description provided by the $q$%
-Gaussian, but also the relative performance of the two distributions when
the comparison is restricted to the asymptotic tail. This allows us to
investigate whether the expected advantage of the power law in the extreme
region is sufficiently pronounced to outweigh the broader statistical
information retained by the $q$-Gaussian, while simultaneously using the
evolution of $q$ to characterize aggregational Gaussianity and the
relaxation between distinct statistical regimes.

The remainder of this paper is organized as follows. In Section \ref%
{Section:Data_Description}, we describe the data employed in this study.
Section \ref{Section:Methodology} presents the methodology used to estimate,
evaluate, and compare the different statistical fits. The results are
presented and discussed in Section \ref{Sec:Results}. Finally, Section \ref%
{Sec:Conclusions} summarizes the main findings and presents our conclusions.

\section{Data Description}

\label{Section:Data_Description}

In this study, we analyze financial data from three distinct groups: U.S.
stocks traded on the New York Stock Exchange (NYSE) and NASDAQ, Brazilian
stocks traded on BOVESPA, and cryptocurrencies. The U.S. stock data span 23
years, from 2003 to 2025; the Brazilian stock data cover the period from
2020 to 2025; and the cryptocurrency data extend over approximately seven
years, from 2017 to 2025.

The assets included in each group were selected to provide a heterogeneous
and representative sample. For the traditional equity markets, both in the
United States and Brazil, we considered companies from a broad range of
economic sectors, including technology, finance, health care, energy,
consumer goods, and industry. The cryptocurrency sample was likewise chosen
to reflect different technological and functional profiles, encompassing
established stores of value, smart-contract platforms, privacy-oriented
currencies, and networks designed for cross-border payments and remittances.
This diversified selection allows us to assess whether the statistical
properties investigated in this work persist across different markets, asset
classes, sectors, and underlying economic or technological structures.

\subsection{Data Collection}

The data used in this study were obtained through the MetaTrader 5 trading
platform, using the market data feed provided by XP Investimentos \footnote{%
https://www.xpi.com.br/}. The entire data-acquisition procedure was
automated in Python through the platform's official integration library. The
resulting financial time series have an intraday resolution of one minute,
allowing us to exploit the longest historical records available on the
broker's server for each market. For the U.S. market, the dataset spans from
2003 to the end of 2025. For the Brazilian market, owing to limitations in
the server-side storage of high-frequency historical data, the available
period extends from late 2020 to the end of 2025.

Cryptocurrency data were obtained from Binance Data Vision \footnote{%
https://data.binance.vision/} , the official public data repository
maintained by the Binance exchange. The acquisition procedure was likewise
fully automated in Python, using programmed requests to retrieve the data
directly from the repository. As for the traditional financial markets,
cryptocurrency prices were sampled at a one-minute intraday resolution. The
corresponding dataset spans from 2017 to the end of 2025.

\section{Methodology for analysis}

\label{Section:Methodology}

\subsection{Returns}

One of the fundamental concepts underlying this study is the financial
return. It represents the relative price difference over a given time
window. One could define this quantity as:

\begin{equation*}
\xi (t)=P(t+\Delta t)-P(t)
\end{equation*}%
where $P(t)$ denotes the asset price. However, many important properties can
be lost with this linear approach, such as time additivity and numerical
symmetry. To address this issue, the most widely used definition for
financial returns is the log-return: 
\begin{equation}
\begin{array}{lll}
\xi (t) & = & \ln (P(t+\Delta t))-\ln (P(t)) \\ 
&  &  \\ 
& \approx & \frac{P(t+\Delta t)}{P(t)}-1%
\end{array}
\label{return}
\end{equation}

\subsection{Power-Laws}

A fundamental aspect in the statistical characterization of financial
markets is the analysis of return distributions. One of their most robust
empirical properties is the occurrence of heavy tails, indicating that
extreme price fluctuations occur considerably more frequently than would be
expected under a Gaussian description \cite{cont2001,gopikrishnan1999,
plerou1999}. A commonly adopted representation for such tails is the pure
power law in $r=|\xi |$,

\begin{equation}
p_{\alpha }(r)=\left\{ 
\begin{array}{ll}
\frac{\alpha -1}{r_{\min }}\left( \frac{r}{r_{\min }}\right) ^{-\alpha }, & 
r\geq r_{\min }, \\[3mm] 
0, & \mathrm{otherwise},%
\end{array}%
\right.  \label{pdf_power_law}
\end{equation}%
where $\alpha $ is the tail exponent and $r_{\min }$ defines the lower bound
above which the power-law behavior is assumed to hold. Power-law tails have
been reported for several financial markets and over a broad range of
temporal scales \cite{mandelbrot1963,fama1965,mantegna1995,lux1996,malevergne2005}. In particular, seminal studies of U.S. financial data found
that the tails of return distributions for market indices and individual
stocks could be described, over substantial ranges, by approximately
inverse-cubic power laws \cite{gopikrishnan1999,plerou1999}. Similar
scale-free behavior has motivated the widespread use of power laws in the
characterization of large financial fluctuations and extreme events \cite%
{gabaix2003}.

Nevertheless, such a description is not expected to remain valid at all
aggregation scales. At relatively short time intervals, $\Delta t$, return
distributions are strongly leptokurtic and heavy-tailed, whereas increasing $%
\Delta t$ generally produces a progressive convergence toward a Gaussian
distribution. This phenomenon, usually referred to as aggregational
Gaussianity, is one of the best-known stylized facts of financial returns 
\cite{cont2001,antypas2013}. Consequently, the range over which a pure power
law provides an adequate representation is intrinsically scale dependent and
may differ substantially among markets and asset classes.

An additional limitation is that a pure power law is intrinsically a model
for the asymptotic tail rather than for the complete return distribution.
Consequently, it does not retain information about the central peak and the
bulk of the distribution, which contain the majority of the observations.
Only data satisfying $|\xi |\ \geq r_{\min }$ contribute directly to the
fitting procedure, making the determination of $r_{\min }$ a central
statistical issue.

The choice of this threshold cannot, in general, be made arbitrarily.
Different values of $r_{\min}$ may lead to different estimates of the
exponent $\alpha$, particularly when the available tail contains only a
limited number of observations. Clauset, Shalizi, and Newman \cite%
{clauset2009} emphasized this difficulty and proposed a systematic framework
based on maximum-likelihood estimation, Kolmogorov--Smirnov statistics, and
likelihood-ratio comparisons for identifying and testing power-law behavior
in empirical data. Even within such a framework, the estimated cutoff and
exponent remain data dependent, and their stability may vary considerably
among different markets, assets, and aggregation scales.

These limitations motivate the search for more general statistical
descriptions capable of representing a broader fraction of the empirical
distribution and of accommodating the changes observed across different
aggregation scales. In particular, it is desirable to employ functional
forms that preserve the heavy-tail behavior observed at short time scales,
while also allowing a smooth evolution toward the Gaussian regime as $\Delta
t$ increases. Such generalized distributions provide the natural framework
for the analysis developed in the following sections.

\subsection{Generalized Exponential}

This function is defined as follows:

\begin{equation}
e_{q}^{x}\equiv \lbrack 1+(1-q)x]^{\frac{1}{1-q}}  \label{q exponencial}
\end{equation}%
The rationale for this formulation comes directly from the limit definition
of the standard exponential function:

\begin{equation}
e^x = \lim_{h \to 0}(1+hx)^{\frac{1}{h}}
\end{equation}
By defining $h = 1-q$, we obtain:

\begin{equation}
e^x = \lim_{q \to 1}[1+(1-q)x]^{\frac{1}{1-q}}
\end{equation}

Thus, we arrive at the generalized exponential, which recovers the standard
exponential in the limit $q\rightarrow 1$. By omitting the limit, we recover
the expression shown in Equation (\ref{q exponencial}). In our work, we
utilize the generalized Gaussian (q-Gaussian) distribution, formulated as:

\begin{equation}
p_{q}(x)=C\ e_{q}^{-\beta x^{2}}=C\ [1-\beta (1-q)x^{2}]^{\frac{1}{1-q}}
\end{equation}%
where $C^{-1}=\int_{-\infty }^{\infty }[1-\beta (1-q)x^{2}]^{\frac{1}{1-q}%
}dx $. The result of this integral depends on the value of the entropic
index $q$. There are three possible regimes:

\begin{enumerate}
\item $q<1$: Eventually, the term $\beta (1-q)x^2$ will exceed one. Given
that $p(x)$ must remain strictly positive, the PDF is defined as zero beyond
this cutoff, which alters the integration limits.

\item $q=1$: In this limit, the q-Gaussian reduces to a standard Gaussian,
yielding $\int_{-\infty}^\infty e^{-\beta x^2} dx = \sqrt{\frac{\pi}{\beta}}$%
.

\item $q>1$: This is the regime of interest for our study (heavy tails). The
term $(1-q)$ becomes negative, so we rewrite the integral using $-(q-1)$.
\end{enumerate}

Following the third case, to solve the integral $\int_{-\infty }^{\infty
}[1+\beta (q-1)x^{2}]^{-\frac{1}{q-1}}dx$, we can draw a direct mathematical
parallel to the probability density function (normalized) of the Student's
t-distribution:

\begin{equation}
f(t)=\frac{\Gamma \left( \frac{\nu +1}{2}\right) }{\sqrt{\nu \pi }\Gamma
\left( \frac{\nu }{2}\right) }\cdot \left( 1+\frac{t^{2}}{\nu }\right) ^{-%
\frac{\nu +1}{2}}
\end{equation}

By matching the exponents, we find the relation for the degrees of freedom $%
\nu =\frac{3-q}{q-1}$, and by matching the arguments of the base, we define
the variable substitution: $t=x\sqrt{\beta (3-q)}$, we conclude that

\begin{equation}
C = \sqrt{\frac{\beta(q-1)}{\pi}} \frac{\Gamma\left(\frac{1}{q-1}\right)}{%
\Gamma\left(\frac{3-q}{2(q-1)}\right)}
\end{equation}

Recognizing that $\frac{3-q}{2(q-1)}=\frac{1}{q-1}-\frac{1}{2}$, we can
rewrite the normalized q-Gaussian Probability Density Function (PDF) as:

\begin{equation}
p_{q}(x)=\frac{\sqrt{\beta (q-1)}\Gamma \left( \frac{1}{q-1}\right) }{\sqrt{%
\pi }\Gamma \left( \frac{1}{q-1}-\frac{1}{2}\right) }[1+\beta (q-1)x^{2}]^{-%
\frac{1}{q-1}},\quad q>1  \label{q_gauss_normalized}
\end{equation}

\subsection{The $k$-th Absolute Moment of the q-Gaussian}

To accurately model financial returns, we analyze the absolute moments of
the distribution. The $k$-th absolute moment $M_k$ is given by integrating $%
|x|^k$ over the entire real line:

\begin{equation}
M_{k}=\langle |x|^{k}\rangle _{q}=2\int_{0}^{\infty }x^{k}p_{q}(x)dx
\end{equation}

Using the known identity for integrals of the form $\int_0^\infty x^{m-1}
(1+ax^2)^{-p} dx$, which evaluates to $\frac{a^{-m/2}}{2} B\left(\frac{m}{2}%
, p - \frac{m}{2}\right)$ (where $B$ is the Beta function), we can
substitute $m = k+1$, $a = \beta(q-1)$, and $p = \frac{1}{q-1}$. This yields:

\begin{equation}
M_{k}=[\beta (q-1)]^{-\frac{k}{2}}\frac{\Gamma \left( \frac{k+1}{2}\right)
\Gamma \left( \frac{1}{q-1}-\frac{k+1}{2}\right) }{\sqrt{\pi }\Gamma \left( 
\frac{1}{q-1}-\frac{1}{2}\right) }  \label{k_moment}
\end{equation}%
Note that this moment only converges if $k<\frac{3-q}{q-1}$.

\subsection{The Method of Moment Ratios}

A major challenge in econophysics is that the absolute moments $M_k$ depend
heavily on the scale parameter $\beta$, which varies drastically across
different assets. To isolate the shape of the tail (dictated solely by the
entropic index $q$), we construct a dimensionless, scale-invariant quantity.

We achieve this by taking the ratio of the $k$-th moment to the $k$-th power
of the first absolute moment ($M_1$). Let us first define $M_1$ by setting $%
k=1$ in Equation (\ref{k_moment}):

\begin{equation}
M_{1}=[\beta (q-1)]^{-\frac{1}{2}}\frac{\Gamma (1)\Gamma \left( \frac{1}{q-1}%
-1\right) }{\sqrt{\pi }\Gamma \left( \frac{1}{q-1}-\frac{1}{2}\right) }
\end{equation}

Raising $M_1$ to the power of $k$ gives:

\begin{equation}
(M_{1})^{k}=[\beta (q-1)]^{-\frac{k}{2}}\left[ \frac{\Gamma (1)\Gamma \left( 
\frac{1}{q-1}-1\right) }{\sqrt{\pi }\Gamma \left( \frac{1}{q-1}-\frac{1}{2}%
\right) }\right] ^{k}
\end{equation}

Now, we define the ratio $R_{q}(k)=\frac{M_{k}}{(M_{1})^{k}}$. Notice how
the scale-dependent term $[\beta (q-1)]^{-\frac{k}{2}}$ appears in both the
numerator and the denominator, canceling out completely and rearranging the
terms, we obtain:

\begin{equation}
R_{q}(k)=\frac{\frac{\Gamma \left( \frac{k+1}{2}\right) \Gamma \left( \frac{1%
}{q-1}-\frac{k+1}{2}\right) }{\sqrt{\pi }\Gamma \left( \frac{1}{q-1}-\frac{1%
}{2}\right) }}{\left[ \frac{\Gamma \left( \frac{1}{q-1}-1\right) }{\sqrt{\pi 
}\Gamma \left( \frac{1}{q-1}-\frac{1}{2}\right) }\right] ^{k}}\text{,}
\label{Eq:Rq}
\end{equation}%
which depends exclusively on $q$ and $k$.

By plotting this theoretical curve against the empirical ratios extracted
from the financial time series, we can optimize the index $q$ using
numerical methods, successfully bypassing the scale parameter estimation.
The most important point here is that $R_{q}(k)$ does not depend on $\beta $.

\subsection{Maximum Likelihood Estimation (MLE) for Power Laws}

While the moment ratio method is exceptionally powerful for mapping the
global shape of the distribution, isolating the extreme tail requires a
different statistical approach. The most robust and widely accepted method
for fitting the tail exponent $\alpha$ of a Power Law is the Maximum
Likelihood Estimation (MLE).

The core principle of MLE is to find the parameter value that maximizes the
probability (likelihood) of observing the given empirical data. Given a set
of $N$ independent observations $x_i \ge x_{min}$ in the heavy tail, the
likelihood function $L(\alpha)$ is the product of the individual
probabilities defined as:

\begin{equation}
L(\alpha )=\prod_{i=1}^{N}p_{\alpha }(x_{i})=\prod_{i=1}^{N}\frac{\alpha -1}{%
x_{\min }}\left( \frac{x_{i}}{x_{\min }}\right) ^{-\alpha }
\end{equation}

Because multiplication of many small probabilities leads to numerical
underflow, it is mathematically more convenient to maximize the natural
logarithm of the likelihood function, denoted as $\mathcal{L}(\alpha)$:

\begin{equation}
\mathcal{L}(\alpha )=\ln L(\alpha )=\sum_{i=1}^{N}\left[ \ln (\alpha -1)-\ln
x_{\min }-\alpha \ln \left( \frac{x_{i}}{x_{\min }}\right) \right]
\end{equation}

Solving $\frac{\partial \mathcal{L}}{\partial \alpha }=0$, finally, we
arrive at the exact analytical estimator for the Power Law tail exponent:

\begin{equation}
\hat{\alpha}=1+N\left[ \sum_{i=1}^{N}\ln \left( \frac{x_{i}}{x_{\min }}%
\right) \right] ^{-1}
\end{equation}

This MLE formula is computationally efficient and statistically consistent.
By applying it to the truncated data ($x\geq x_{\min }$), we extract the
pure tail behavior, which can then be formally compared to the asymptotic
limit of the non-extensive q-Gaussian model.

\subsection{Estimating the Optimal Tail Cutoff ($x_{\min }$)}

The Maximum Likelihood Estimator (MLE) accurately dictates the scaling
parameter $\alpha $ given a specific starting point of the tail, $x_{\min }$%
. However, if $x_{\min }$ is chosen too low, the power law model will be
forcibly fitted to the body of the distribution, significantly biasing the
exponent. Conversely, if $x_{\min }$ is set too high, valuable empirical
data is discarded, drastically increasing the statistical uncertainty of the
estimate.

To rigorously define the boundary where the Gaussian-like body ends and the
heavy tail begins, we employ the Kolmogorov-Smirnov (KS) statistic. The KS
test quantifies the maximum absolute distance between the empirical
Complementary Cumulative Distribution Function (CCDF) of the data and the
theoretical CCDF of the fitted model. For the empirical data confined to $%
x\geq x_{\min }$, the empirical CCDF $S(x)$ is defined as the fraction of
data points with a value greater than or equal to $x$, i.e., given a random sample $x_1, x_2, \dots, x_N$ of size $N$, the function $S(x)$ represents the proportion of observations that are greater than or equal to a real value $x$. Formally, this relationship is expressed using the cardinality of the subset of observations that satisfy the condition, as shown in Equation \eqref{eq:proportion}:

\begin{equation}
\label{eq:proportion}
S(x) = \frac{\left| \{ i \in \{1, \dots, N\} : x_i \ge x \} \right|}{N}
\end{equation}
where $\lvert \cdot \rvert$ denotes the cardinality (or number of elements) of the set.

For the theoretical Power Law model, the CCDF $F(x)$ represents the
probability that a value is greater than or equal to $x$, obtained by
integrating the probability density function from $x$ to infinity:

\begin{equation}
F(x)=\int_{x}^{\infty }p_{\alpha }(r)dr=\int_{x}^{\infty }\frac{\alpha -1}{%
x_{\min }}\left( \frac{r}{x_{\min }}\right) ^{-\alpha }dr
\end{equation}

Evaluating this integral yields the theoretical CCDF:

\begin{equation}
F(x)=\left( \frac{x}{x_{\min }}\right) ^{1-\alpha }
\end{equation}

The Kolmogorov-Smirnov distance $D$ is defined as the supremum of the
absolute difference between the empirical and theoretical distributions
across all $x \ge x_{min}$:

\begin{equation}
D = \max_{x \ge x_{min}} | S(x) - F(x) |
\end{equation}

The optimization procedure dictates that we must iterate over every unique
value in the dataset, treating each as a candidate for $x_{min}$. For each
candidate, we compute the corresponding MLE $\hat{\alpha}$ and then
calculate the KS distance $D$. The optimal cutoff $x_{min}^{\ast }$ is
determined as the value that minimizes $D$, thereby ensuring the empirical
tail follows the theoretical power law model as closely as possible:

\begin{equation}
x_{min}^* = \arg \min_{x_{min}} \left( \max_{x \ge x_{min}} | S(x) - F(x) |
\right)
\end{equation}

\subsubsection{Convergence to the Asymptotic Relation}

Once isolated to the extreme tail ($x>>1$), the constant $1$ in the
q-Gaussian base becomes negligible: 
\begin{equation}
\lbrack 1+(q-1)\beta x^{2}]^{-\frac{1}{q-1}}\approx \left[ (q-1)\beta x^{2}%
\right] ^{-\frac{1}{q-1}}\propto x^{-\frac{2}{q-1}}
\end{equation}

Because the Power Law decays as $x^{-\alpha}$, matching the asymptotic
behavior of the functions requires matching their exponents directly: 
\begin{equation}
\alpha = \frac{2}{q-1}
\end{equation}

The properly truncated moment curves collapse exactly onto this asymptotic
relation, proving that in the extreme tail, the non-extensive q-Gaussian and
the Power Law are mathematically indistinguishable.

\subsection{The Fitting Procedure:}

To fit empirical financial data, considering that we have $N$ values of
returns $\xi _{1},$ $\xi _{2},...,\xi _{N}$, we compute the empirical ratios:%
\begin{equation*}
R(k)=\frac{N^{k-1}\sum_{i=1}^{N}\left\vert \xi _{i}\right\vert ^{k}}{\left(
\sum_{i=1}^{N}\left\vert \xi _{i}\right\vert \right) ^{k}}
\end{equation*}
across a fractional array of $k$ values. We then minimize the residuals
between these empirical values and the theoretical function $R_{q}(k)$ using
a non-linear least squares optimization, where: 
\begin{equation}
q_{opt}=\arg \min_{q\in Q}\frac{1}{N_{k}}\sum_{k\in \mathcal{K}%
}(R_{q}(k)-R(k))^{2}
\end{equation}%
where $Q$ is a set of possible values of $q$ while the sum is made over a
set of $k-$values: $\mathcal{K}=\{k_{1},k_{2},...,k_{N_{k}}\}$.

\subsection{Statistics for Assessing Goodness of Fit}

To assess the analytical fit of the q-Gaussian and power-law models to
empirical financial data, we employ a two-layer statistical approach. The
first layer consists of empirical distribution function (EDF) statistics,
which quantify the geometric distance between the empirical data and the
theoretical probability distributions. The second layer uses a
likelihood-based model-selection framework to establish the statistical
significance of the observed differences. This comprehensive methodology
ensures that the assessment captures both global adherence and accuracy in
the extreme tails.

\subsubsection{Empirical Distribution Function (EDF) Statistics}

Let $x_{1},x_{2},\dots ,x_{N}$ be an independent and identically distributed
sample extracted from a financial-return time series and sorted in
increasing order. The empirical cumulative distribution function (ECDF),
denoted by $F_{N}(x)$, represents the proportion of observations less than
or equal to $x$ and is mathematically defined as $F_{N}(x)=\frac{1}{N}%
\sum_{i=1}^{N}\mathbf{1}_{\{x_{i}\leq x\}}$. We compare this empirical step
function with the continuous cumulative distribution function (CDF) of the
theoretical model, $F(x)$ \footnote{%
It is important to note that, unlike in the analysis used to estimate the
optimal tail cutoff, here we employ the cumulative distribution function
(CDF) rather than the complementary cumulative distribution function (CCDF).
This choice does not affect the numerical results.}. To assess the goodness
of fit rigorously, we employ three distinct metrics, each with a specific
structural sensitivity. For their computational evaluation, we use the
probability integral transform and define $z_{i}=F(x_{i})$ for the ordered
empirical data \footnote{We compute these goodness-of-fit metrics for the q-Gaussian distribution over both the full data domain and the tail region, and for the pure power law exclusively in the tail.When the analysis is restricted to the tail region ($x \geq x_{min}$), the theoretical probabilities must be conditioned on this survival threshold to prevent scaling mismatches against the truncated empirical data. In this regime, we substitute $F(x)$ with the conditional CDF $F_{cond}(x) = 1 - \frac{1 - F(x)}{1 - F(x_{min})}$.}.

\subsubsection{Kolmogorov--Smirnov (KS) Statistic}

The Kolmogorov--Smirnov (KS) statistic is a nonparametric metric \cite{kolmogorov1933} that
evaluates the supremum---the maximum absolute vertical distance---between
the empirical distribution and the theoretical cumulative distribution. It
is theoretically defined as:

\begin{equation}
D = \sup_x |F_N(x) - F(x)|
\end{equation}

For a discrete computational implementation, this distance is evaluated at
each empirical data point, ensuring that both the upper and lower edges of
the empirical step are compared with the theoretical curve:

\begin{equation}
D = \max_{1 \le i \le N} \left[ \max \left( \frac{i}{N} - z_i, z_i - \frac{%
i-1}{N} \right) \right]
\end{equation}

Intuitively, the KS statistic identifies the single point of greatest
divergence between the model and the empirical data, yielding a
dimensionless probability error strictly bounded between 0 and 1. Because it
evaluates only the maximum absolute deviation, it provides a strict
worst-case assessment of the global fit. Its sensitivity, however, is
generally concentrated near the median of the distribution, making it
relatively insensitive to deviations in the asymptotic tails.

\subsubsection{The Cram\'{e}r--von Mises Test Statistic ($W^{2}$)}

Whereas the KS statistic isolates a single maximum error, the Cram\'{e}%
r--von Mises (CvM) statistic \cite{cramer1928} assesses the global discrepancy by integrating
the squared deviations over the entire continuous distribution, scaled by
the sample size $N$:

\begin{equation}
W^2 = N \int_{-\infty}^{\infty} (F_N(x) - F(x))^2 \, dF(x)
\end{equation}

In algorithmic practice, evaluating this integral over the transformed
uniform variables $z_i$ yields the following operational sum:

\begin{equation}
W^2 = \frac{1}{12N} + \sum_{i=1}^N \left( z_i - \frac{2i - 1}{2N} \right)^2
\end{equation}

The CvM statistic acts as a continuous score of global geometric divergence.
By squaring and summing the errors, it inherently assigns greater
mathematical weight to regions containing most of the empirical
observations. Consequently, this metric is highly sensitive to the body (the
mass) of the distribution and strongly penalizes models that exhibit
systematic geometric deviations across broad regions of ordinary data.

\subsubsection{Anderson--Darling (AD) Statistic}

The Anderson--Darling (AD) statistic \cite{anderson1952} is a structural refinement of the CvM
metric, specifically designed to increase sensitivity near the boundaries of
the distribution. This is achieved by introducing a logarithmic weighting
function in the denominator that disproportionately penalizes discrepancies
in the extreme quantiles:

\begin{equation}
A^2 = N \int_{-\infty}^{\infty} \frac{(F_N(x) - F(x))^2}{F(x)(1 - F(x))} \,
dF(x)
\end{equation}

Computationally, the discrete sum used to calculate the AD statistic
accurately while avoiding floating-point \textit{underflow} at the
boundaries is given by:

\begin{equation}
A^{2}=-N-\frac{1}{N}\sum_{i=1}^{N}(2i-1)\left[ \ln (z_{i})+\ln (1-z_{N-i+1})%
\right]
\end{equation}

The theoretical weighting factor diverges as $F(x)\to0$ or $F(x)\to1$.
Because of this mathematical amplification, even small predictive errors for
the rarest market events (for example, severe financial crashes located deep
in the tails) lead to a substantial increase in the AD error score. In
econophysics, the AD statistic therefore acts as a strict tail inspector,
ensuring that a model is judged by its ability to characterize extreme,
scale-invariant asymptotic behavior rather than merely fitting the central
noise.

\subsection{Vuong Closeness Test (Log-Likelihood Ratio)}

EDF statistics such as KS, CvM, and AD provide an absolute measure of
geometric distance, but they cannot inherently determine whether the
difference in fit between two competing models is statistically significant.
To address this issue, we employ the Vuong closeness test, a
likelihood-ratio-based method \cite{vuong1989}.

In our methodology, we compare a specialized model fitted exclusively to
extreme events (the power law) with a unified model that captures the full
market dynamics (the q-Gaussian). To ensure a fair and rigorous comparison,
the Vuong test is evaluated strictly within the truncated tail domain ($%
|x|\ge x_{min}$).

For each empirical observation $x_i$ in the tail, we calculate the pointwise
log-likelihood ratio ($\mathcal{R}_i$) between the conditional probability
densities of the q-Gaussian ($P_{qG}$) and the power law ($P_{PL}$):

\begin{equation}
\mathcal{R}_i = \ln P_{qG}(x_i) - \ln P_{PL}(x_i)
\end{equation}

The test operates under the null hypothesis that both models are equally
close to the true data distribution. The test statistic $V$ is calculated by
standardizing the mean of these pointwise likelihood ratios ($\bar{\mathcal{R%
}}$) by their empirical standard deviation ($\omega$), rescaled by the
number of observations in the tail ($N_{tail}$):

\begin{equation}
V = \frac{\sqrt{N_{tail}} \, \bar{\mathcal{R}}}{\omega}
\end{equation}

Under general conditions, the statistic $V$ follows a standard normal
distribution, $\mathcal{N}(0,1)$. Intuitively, it measures which model
assigned a higher theoretical probability to the extreme events that
actually occurred in the empirical data.

The statistical decision at the standard 95\% confidence level is
straightforward:

\begin{itemize}
\item $V>1.96$ ($p<0.05$)\textbf{:} The q-Gaussian is significantly superior
in the tail regime.

\item $V<-1.96$ ($p<0.05$)\textbf{:} The power law is significantly superior
in the tail regime.

\item $-1.96\leq V\leq 1.96$\textbf{:} A statistical tie. The available tail
data are insufficient to establish conclusively that the predictive power of
one model exceeds that of the other.
\end{itemize}

By combining the geometric rigor of EDF statistics with the Vuong likelihood
framework, this approach ensures that any claim of model superiority in
extreme regimes is rigorously validated against the inherent noise of
financial markets.

\subsection{Nonparametric Bootstrap Resampling for Goodness-of-Fit Metrics}

To assess the comparative robustness of the q-Gaussian and the power law,
and to determine whether the nominal superiority of one model over the other
in the error metrics is statistically significant, we implement a
nonparametric \textit{bootstrap} resampling protocol \cite{efron1979}. The methodological
objective of this stage is not to re-estimate parameters, but to isolate
purely sampling variance in the extreme-tail region and observe how it
affects the geometric distance between the data and the already optimized
theoretical distributions.

The computational procedure was structured as follows:

\begin{enumerate}
\item \textbf{Fixing the Theoretical Parameters:} The globally optimal
q-Gaussian parameters ($q$ and $\beta $) and the local power-law parameters (%
$\alpha $ and $x_{min}$) are held fixed after being fitted previously. We
isolate the vector of returns composing the extreme tail ($|x|\ \geq x_{min}$%
).

\item \textbf{Generation of Synthetic Samples:} From this empirical tail of
size $N_{tail}$, we generate a new synthetic sample by randomly drawing $%
N_{tail}$ observations with replacement. This procedure is repeated to
create $B=2500$ independent \textit{bootstrap} samples for each asset and
each time scale.

\item \textbf{Calculation of Error Differences:} For each generated
synthetic sample, we calculate the three discrete empirical-distribution
statistics (KS, CvM, and AD) relative to both theoretical models. We then
obtain the performance difference for each metric, defined as $\Delta \text{%
Error}=\text{Error}_{qG}-\text{Error}_{PL}$.

\item \textbf{Confidence Intervals and Decision Criterion:} After the $2500$
iterations, we construct an empirical 95\% confidence interval for the error
differences by evaluating the 2.5\% and 97.5\% percentiles. The adopted
criterion for statistical superiority is:

\begin{itemize}
\item If the lower bound of the interval is strictly greater than zero ($%
\Delta \text{Error}>0$), the power law exhibits a consistently smaller error
in 95\% of the scenarios and is declared significantly superior.

\item If the upper bound of the interval is strictly less than zero ($\Delta 
\text{Error}<0$), the q-Gaussian is declared statistically superior.

\item If the confidence interval contains zero, a \textbf{statistical tie}
is declared. This shows that the natural sampling variability of the tail is
greater than the actual performance difference between the models, rendering
them statistically indistinguishable.
\end{itemize}
\end{enumerate}

This approach ensures that any claim of geometric superiority is not merely
an artifact of isolated fluctuations in a finite sample, establishing a
rigorous mathematical significance threshold for the direct comparison of
deep-tail fits.

\section{Results and Discussion}

\label{Sec:Results}

In this section, we present our main results, which are divided into two
parts: stock markets and cryptocurrencies.

\subsection{Stock-Market Data}

\begin{figure}[tbph]
\centering
\par
\begin{subfigure}{0.48\textwidth}
        \centering
        \includegraphics[width=\linewidth]{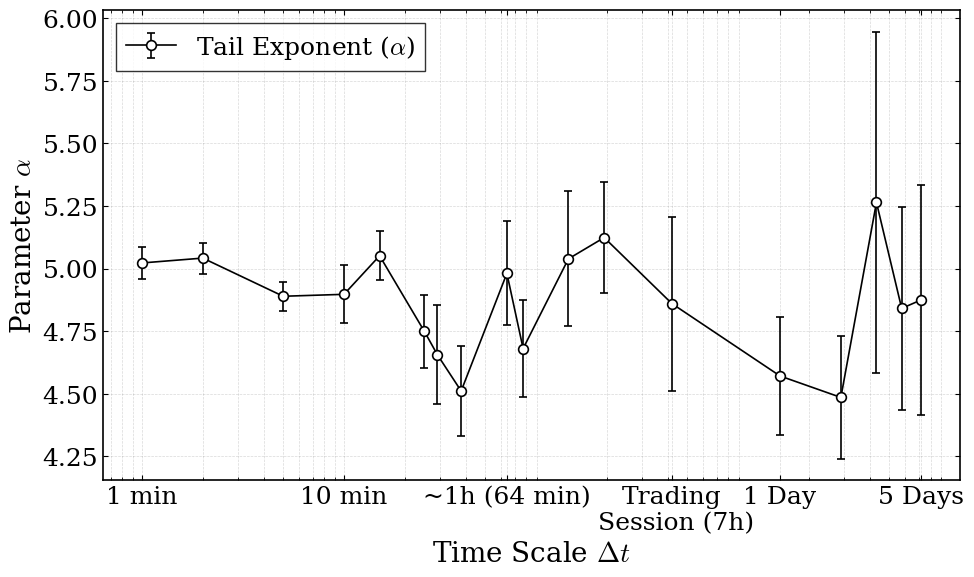}
        \caption{Behavior of the Power-Law Exponent ($\alpha$) -- Brazilian Stocks.}
        \label{fig:esq1}
    \end{subfigure}\hfill 
\begin{subfigure}{0.48\textwidth}
        \centering
        \includegraphics[width=\linewidth]{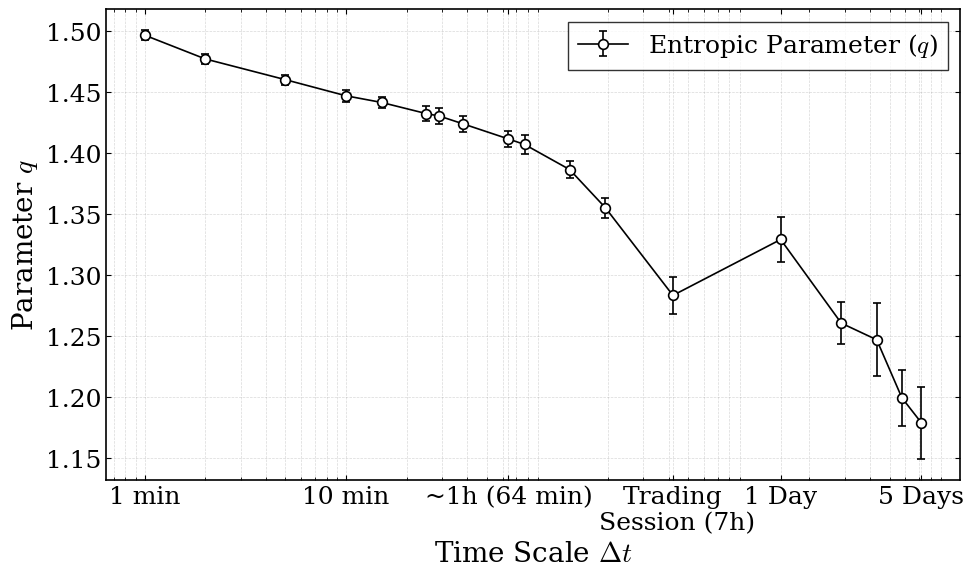}
        \caption{Behavior of the $q$-Gaussian Parameter ($q$) -- Brazilian Stocks.}
        \label{fig:dir1}
    \end{subfigure}
\par
\vspace{0.5cm} 
\par
\begin{subfigure}{0.48\textwidth}
        \centering
        \includegraphics[width=\linewidth]{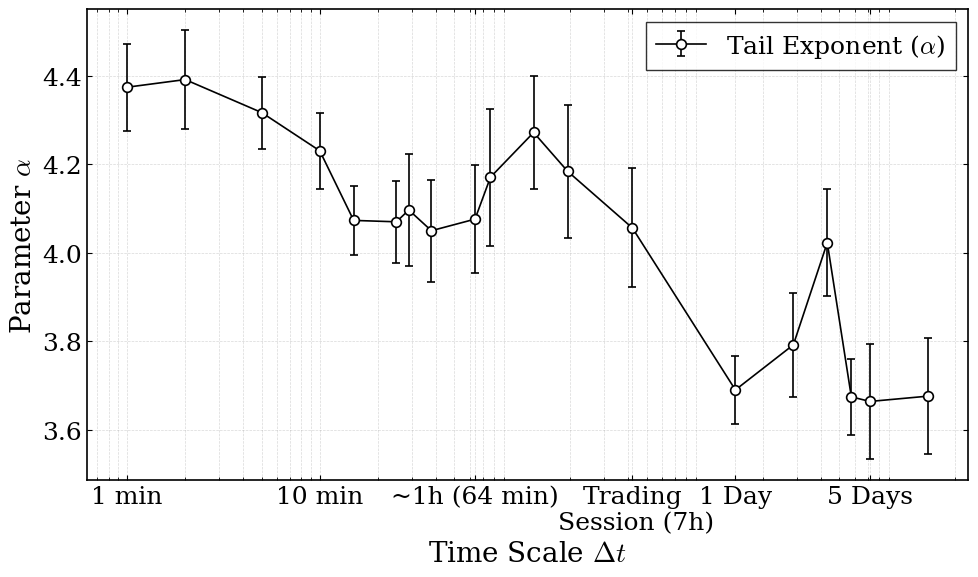}
        \caption{Behavior of the Power-Law Exponent ($\alpha$) -- U.S. Stocks.}
        \label{fig:esq2}
    \end{subfigure}\hfill 
\begin{subfigure}{0.48\textwidth}
        \centering
        \includegraphics[width=\linewidth]{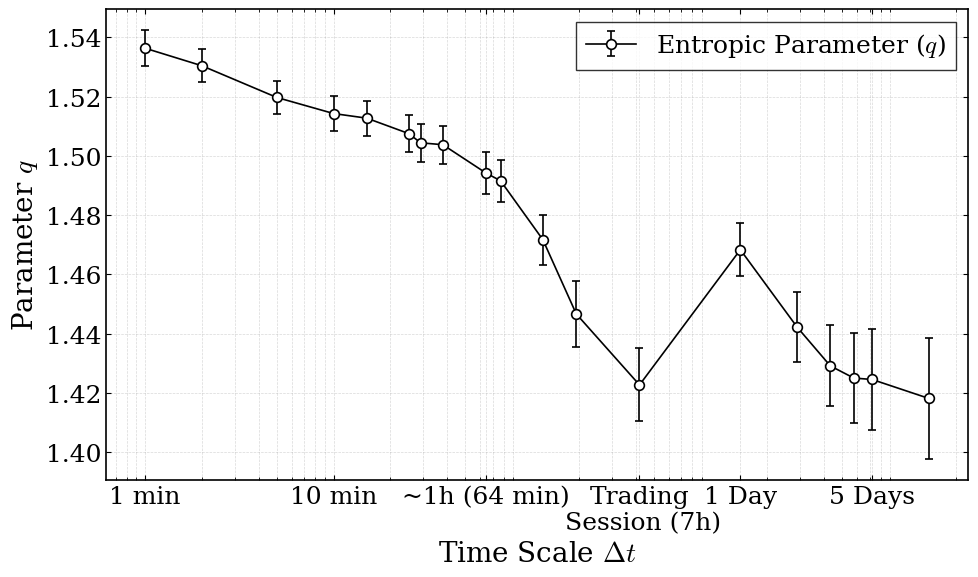}
        \caption{Behavior of the $q$-Gaussian Parameter ($q$) -- U.S. Stocks.}
        \label{fig:dir2}
    \end{subfigure}
\caption{Behavior of $\alpha$ and $q$ as functions of the log-return time scale. 
The uncertainties correspond to the standard errors of the mean across the 21 
stocks analyzed in the U.S. market and the 15 stocks analyzed in the Brazilian 
market.}
\label{fig:comparacao_grid}
\end{figure}

The stock-market results first reveal a limitation of the power-law fit to
capture aggregational Gaussianity. In both stock markets analyzed, the
values of $\alpha $ obtained using the method proposed by Clauset, Shalizi,
and Newman \cite{clauset2009} remain relatively stable as the return
interval increases, contrary to what is expected from the theory of
aggregational Gaussianity. From a theoretical standpoint, the exponential
decay characteristic of a Gaussian distribution would imply that the
power-law exponent diverges to infinity ($\alpha \rightarrow \infty $). This
behavior arises because the fitting procedure will always attempt to find a
power law that matches the data, even when the presence of a genuine power
law is no longer supported. Indeed, by examining the behavior of $\alpha $
for U.S. stocks, the reader might incorrectly infer that market volatility
increases with the time scale because of the downward trend of the
parameter, a conclusion that is entirely inconsistent with the expected
behavior of financial returns. By contrast, the $q$-Gaussian fit captures
aggregational Gaussianity remarkably well, as shown in Fig.~\ref%
{fig:comparacao_grid}. We can see four subfigures: Fig.~\ref%
{fig:comparacao_grid} (a), which shows the behavior of the power-law
exponent $\alpha $ as a function of the time scale $\Delta t$, Fig.~\ref%
{fig:comparacao_grid} (b), showing the behavior of the parameter $q$ as a
function of the time scale $\Delta t$ for Brazilian stocks. Similarly, Figs.~%
\ref{fig:comparacao_grid} (c) and (d) show the behavior of the same
parameters, respectively, for U.S. stocks.

Not only is the behavior of the fits more coherent and consistent with
theory and observation, but the uncertainty associated with the points\ --
in this case, the sample dispersion -- is also considerably smaller for the $%
q$--Gaussian in comparison with the power-law fit. The main reason is that
fitting the global parameter $q$ uses all the data available at each time
scale, whereas the power-law fit selects only the most extreme observations.
The principal difficulty faced by the power law is that, as the time
interval increases, the data become increasingly scarce and the tails
progressively shorter, so the fit can no longer capture the system behavior
reliably. This produces unstable results for small samples and large
asset-to-asset differences. The global nature of the $q$-Gaussian fit
greatly mitigates this problem.

This discrepancy is also explained by the mathematical nature of the
power-law fitting protocol itself. The standard and statistically robust
methodology for extreme-tail modeling (again, following \cite{clauset2009})
does not assess the distribution as a whole; instead, it employs a two-step
procedure: MLE to determine the exponent $\alpha $, coupled with
minimization of the KS distance to identify the optimal cutoff $x_{min}$.

By minimizing the KS statistic strictly within the extreme region ($|x|\
\geq x_{min}$), the algorithm is designed to isolate the subsample that best
fits an algebraic density function, thereby mathematically separating it
from the behavior of most of the distribution. Consequently, even when the
empirical data lose the strict power-law property and the system as a whole
converges toward the Gaussian regime, the domain restriction and KS
minimization criterion force a local power-law fit over the few remaining
deep-tail events. The result is a spurious finite exponent. The fitting
procedure alone is therefore insufficient. In addition to estimating the
parameters, further statistical tests are required to verify whether the
power-law hypothesis actually provides an adequate representation of the
dataset.

\begin{figure}[tbph]
\centering
\begin{subfigure}{0.32\textwidth}
        \includegraphics[width=\linewidth]{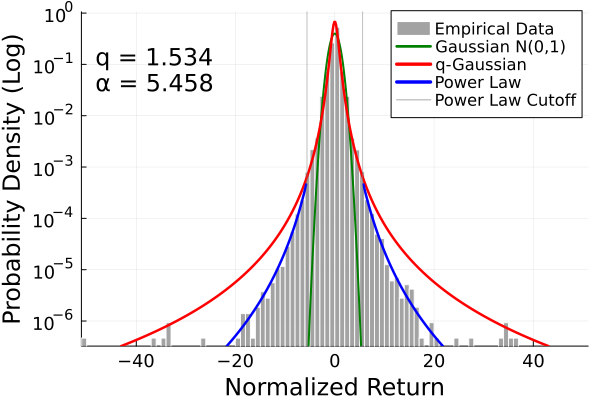}
        \caption{$\Delta t = 1$ min.}
    \end{subfigure}\hfill 
\begin{subfigure}{0.32\textwidth}
        \includegraphics[width=\linewidth]{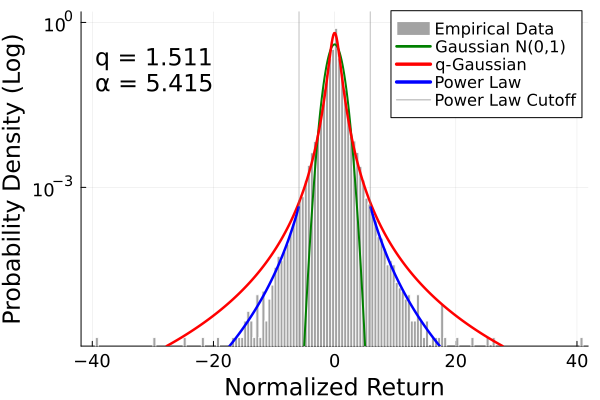}
        \caption{$\Delta t = 2$ min.}
    \end{subfigure}\hfill 
\begin{subfigure}{0.32\textwidth}
        \includegraphics[width=\linewidth]{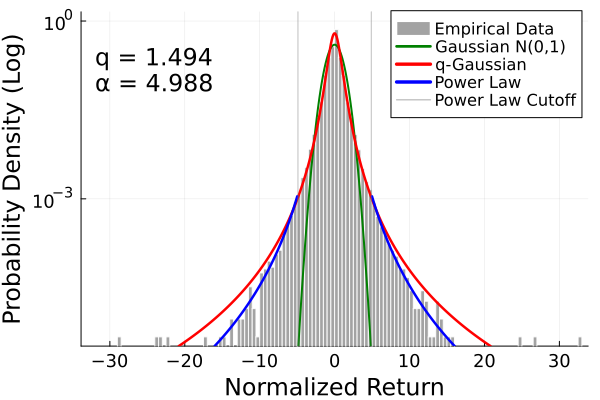}
        \caption{$\Delta t = 5$ min.}
    \end{subfigure}
\par
\vspace{0.3cm} 
\begin{subfigure}{0.32\textwidth}
        \includegraphics[width=\linewidth]{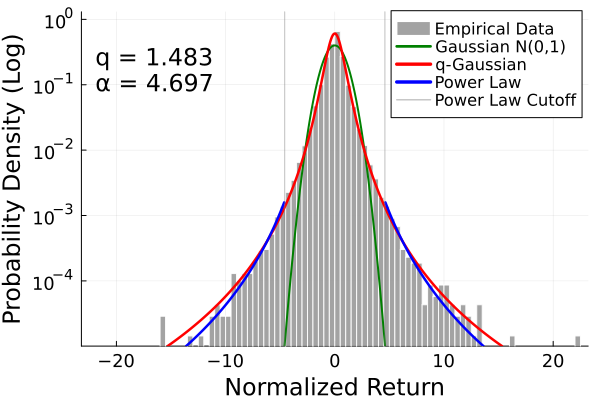}
        \caption{$\Delta t = 15$ min.}
    \end{subfigure}\hfill 
\begin{subfigure}{0.32\textwidth}
        \includegraphics[width=\linewidth]{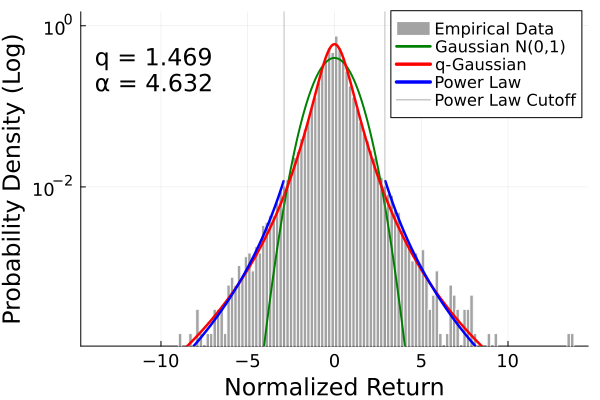}
        \caption{$\Delta t = 64$ min.}
    \end{subfigure}\hfill 
\begin{subfigure}{0.32\textwidth}
        \includegraphics[width=\linewidth]{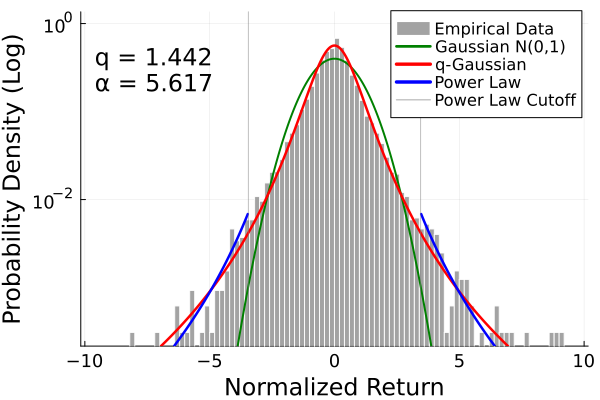}
        \caption{$\Delta t = 129$ min.}
    \end{subfigure}
\par
\vspace{0.3cm} 
\begin{subfigure}{0.32\textwidth}
        \includegraphics[width=\linewidth]{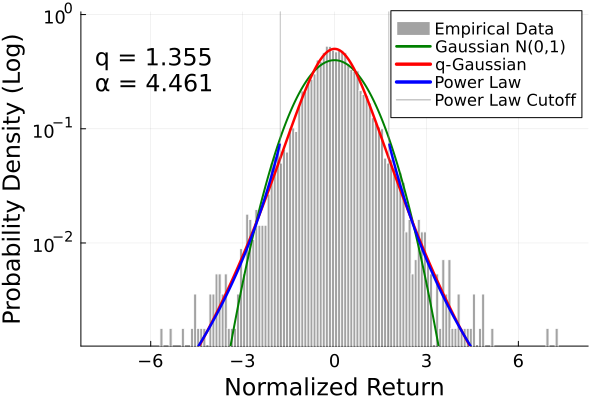}
        \caption{Full Intraday Period.}
    \end{subfigure}\hfill 
\begin{subfigure}{0.32\textwidth}
        \includegraphics[width=\linewidth]{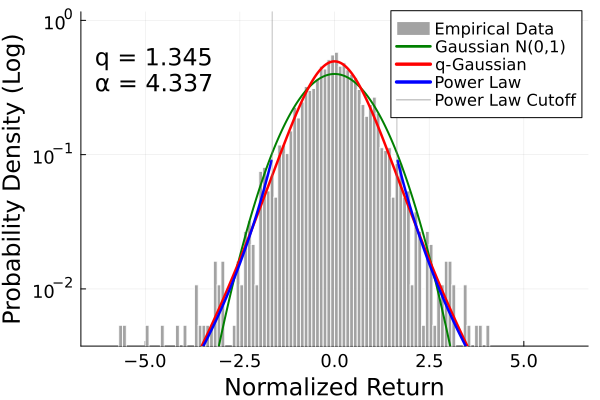}
        \caption{3 Days.}
    \end{subfigure}\hfill 
\begin{subfigure}{0.32\textwidth}
        \includegraphics[width=\linewidth]{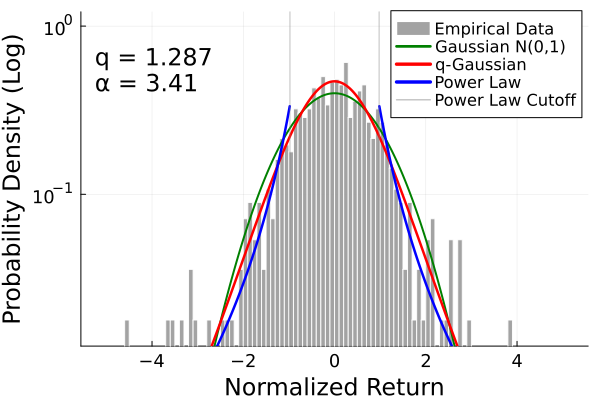}
        \caption{10 Days.}
    \end{subfigure}
\caption{Behavior of the empirical return distributions and the
corresponding global and local theoretical fits for AMD stock across
multiple aggregation scales\ ($\Delta t$).}
\label{fig:amd_painel_distribuicoes}
\end{figure}

Taking the historical evolution of AMD (Advanced Micro Devices) as an
example in Fig.~\ref{fig:amd_painel_distribuicoes}, the results indicate
that evaluating extreme events exclusively through this local-minimization
protocol can obscure the macroscopic statistical behavior of the system. The
persistence of finite $\alpha $ values may incorrectly suggest the survival
of heavy tails, whereas the global parameter $q$ indicates that the
distribution is already transitioning toward the Gaussian regime. This
observation provides a methodological justification for using nonextensive
probability-density functions to characterize financial dynamics
comprehensively across different temporal resolutions.

A direct comparison of the two stock markets also reveals a quantitative
difference in the magnitude of extreme fluctuations, accurately captured by
the Tsallis parameter. U.S. assets systematically exhibit higher values of
the entropic parameter $q$ than Brazilian assets at intraday scales,
reflecting a trading environment with substantially heavier tails in the
United States. From an econophysics perspective, this difference in the
level of $q$ may be attributed to market microstructure: the enormous
liquidity, predominance of ultra-high-frequency algorithmic trading (HFT),
and informational complexity of U.S. exchanges create a system that is more
prone to very-short-term extreme shocks. The Brazilian market, in turn,
exhibits slightly lower values of $q$, indicating dynamics that, although
still anomalous and non-Gaussian, are quantitatively less extreme and
converge toward Gaussian behavior much more rapidly.

\subsubsection{The Overnight Anomaly in Parameter Convergence}

An interesting feature of the parameter evolution in the stock market is
that, although the entropic index $q$ exhibits an overall decay toward the
Gaussian boundary ($q\rightarrow 1$) because of temporal aggregation
governed by the Central Limit Theorem, a distinct localized deviation occurs
precisely at the one-day scale (\textquotedblleft 1\ Day\textquotedblright
). This statistical anomaly is mathematically associated with the inclusion,
from this point onward, of the interval between trading days -- the
overnight effect. The interruption of continuous trading, combined with the
accumulation of latent macroeconomic information while the market is closed,
generates substantial price jumps at the subsequent opening \cite{french1980,wood1985,mcinishwood1985,harris1986}. For intervals
shorter than one day, these returns were removed to remain faithful to the
definition of intraday returns; from the one-day scale onward, however,
returns were calculated from differences between the closing prices of
consecutive days.

To quantify this phenomenon explicitly, a separate analysis was performed
using strictly \textit{overnight} returns, measured between the previous
day's closing price and the current day's opening price. The empirical
estimate for this specific subset reveals an exceptionally heavy-tailed
distribution, yielding a mean power-law exponent of $\alpha =3.133\pm 0.062$
and a mean $q$-Gaussian entropic index of $q=1.548\pm 0.006$ across the
analyzed U.S. assets. These values indicate a severe departure from
normality, confirming that information shocks outside trading hours strongly
inflate the probability mass in the extreme quantiles. Consequently, when
intraday data are aggregated at daily resolutions, the empirical
distribution inherits this localized increase in nonextensivity, causing a
temporary peak in $q$ and a corresponding decrease in $\alpha $ before
multi-day temporal aggregation resumes its convergence toward the normal
characteristics.

\subsubsection{Empirical Goodness of Fit}

Tracking the behavior of the goodness-of-fit metrics reveals a consistent
phenomenological pattern across all three test statistics --
Kolmogorov--Smirnov, Cram\'{e}r--von Mises, and Anderson--Darling -- for
both Brazilian and U.S. stocks, as shown in Fig. \ref%
{fig:comparacao_6_imagens}. We divide the analysis into three different
fitting procedures: the $q$--Gaussian fitted to the entire dataset (global
fit), the $q$--Gaussian fitted to the tail, and the power-law fit to the
tail. The global $q$--Gaussian fit improves continuously, as indicated by
the decreasing error, with increasing time scale $\Delta t$, reaching its
best agreement near the one-day scale before tending to stabilize. This
behavior reflects the increasing ability of the model to describe the return
distribution as high-frequency microstructural effects are progressively
suppressed under temporal aggregation. At the same time, the performance of
the$\ q$--Gaussian restricted to the tail remains relatively stable, with
modest improvements at intermediate time scales. By contrast, the quality of
the power-law fit deteriorates monotonically under temporal aggregation.
Nevertheless, despite this continuous degradation, the absolute error
associated with the power-law fit remains systematically smaller than that
of the $q$--Gaussian tail fit over all the time scales considered.

\begin{figure}[tbph]
    \centering
    
    \begin{subfigure}[t]{0.48\textwidth}
        \centering
        \includegraphics[width=\linewidth]{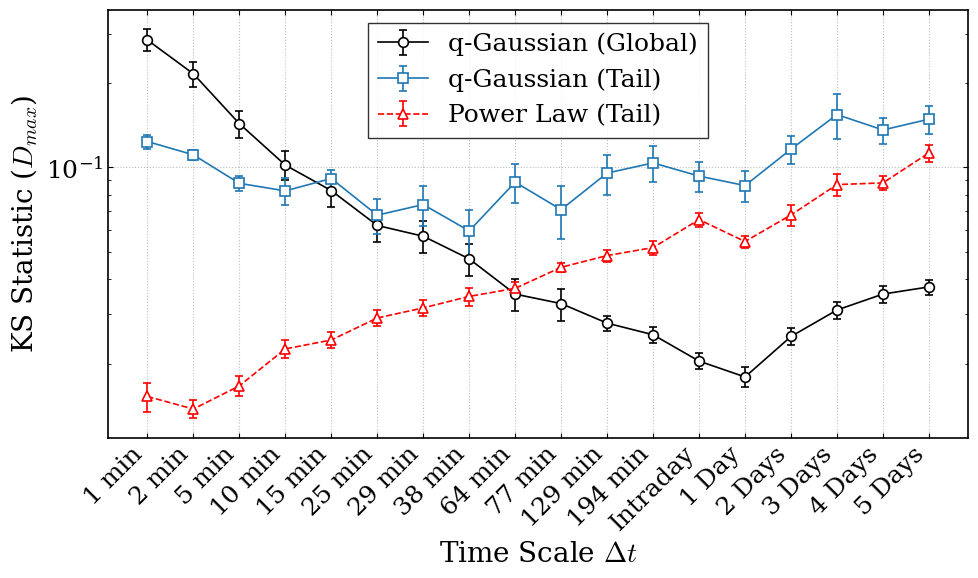}
        \caption{Behavior of the Kolmogorov--Smirnov Distance ($D_{\max}$) -- Brazil.}
        \label{fig:esq1}
    \end{subfigure}
    \hfill
    \begin{subfigure}[t]{0.48\textwidth}
        \centering
        \includegraphics[width=\linewidth]{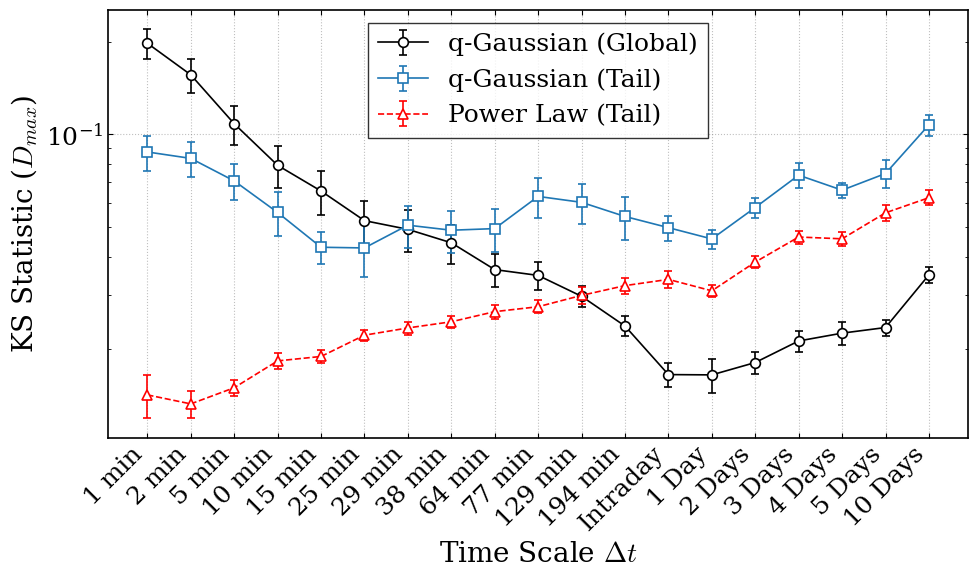}
        \caption{Behavior of the Kolmogorov--Smirnov Distance ($D_{\max}$) -- U.S.}
        \label{fig:dir1}
    \end{subfigure}
    
    \vspace{0.3cm}
    
    \begin{subfigure}[t]{0.48\textwidth}
        \centering
        \includegraphics[width=\linewidth]{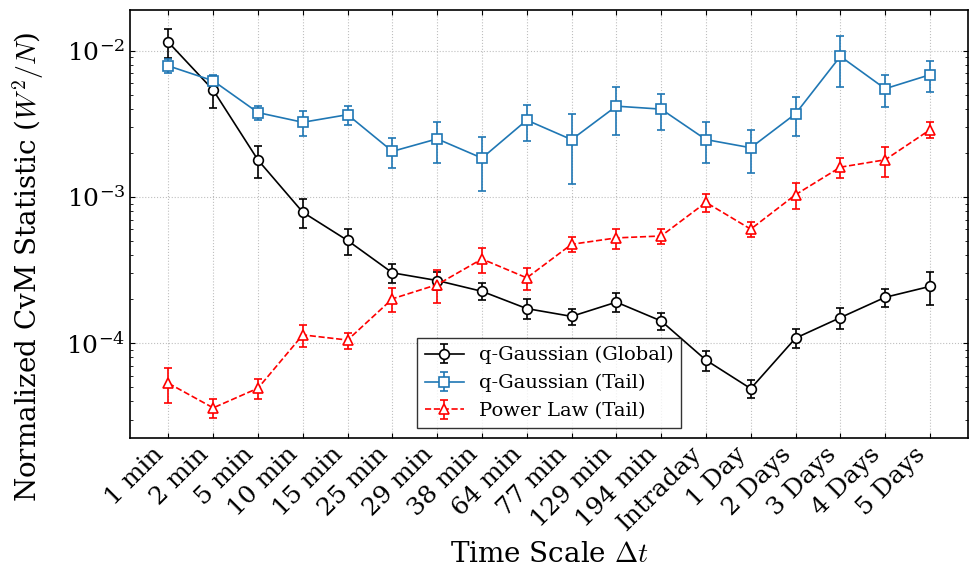}
        \caption{Behavior of the Normalized Cramer--von Mises Statistic ($W^2/N$) -- Brazil.}
        \label{fig:esq2}
    \end{subfigure}%
    \hfill
    \begin{subfigure}[t]{0.48\textwidth}
        \centering
        \includegraphics[width=\linewidth]{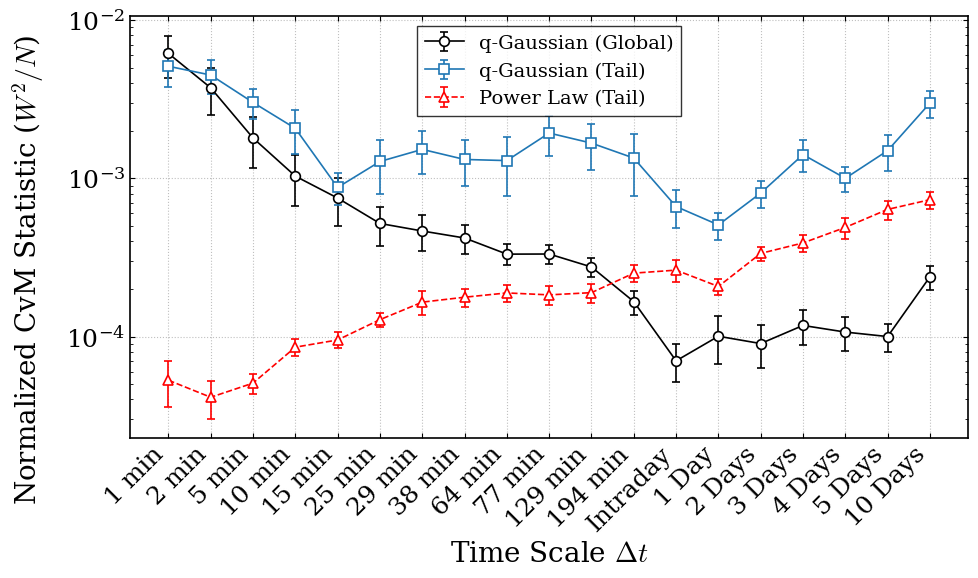}
        \caption{Behavior of the Normalized Cramer--von Mises Statistic ($W^2/N$) -- U.S.}
        \label{fig:dir2}
    \end{subfigure}
    
    \vspace{0.3cm}
    
    \begin{subfigure}[t]{0.48\textwidth}
        \centering
        \includegraphics[width=\linewidth]{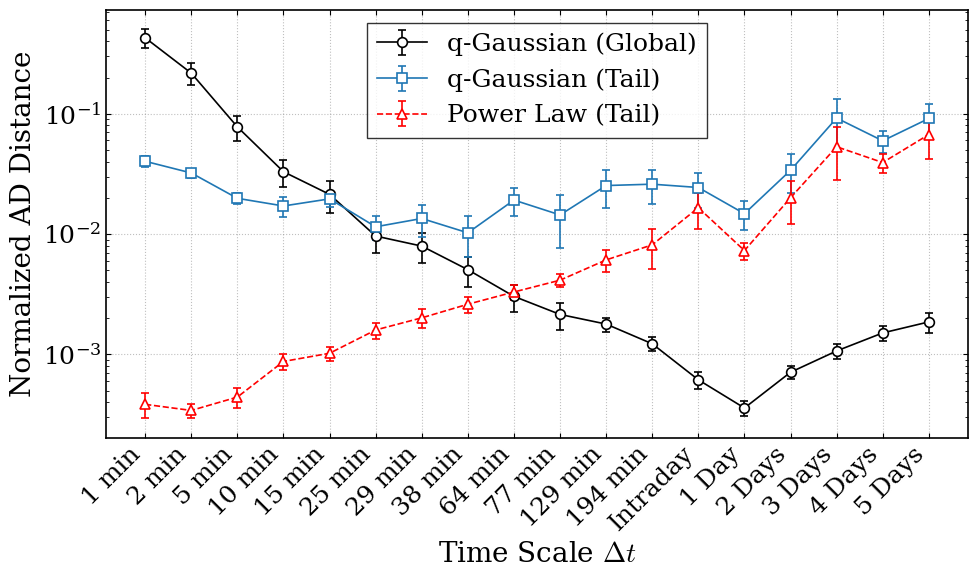}
        \caption{Behavior of the Normalized Anderson--Darling Statistic ($A^2/N$) -- Brazil.}
        \label{fig:esq3}
    \end{subfigure}%
    \hfill
    \begin{subfigure}[t]{0.48\textwidth}
        \centering
        \includegraphics[width=\linewidth]{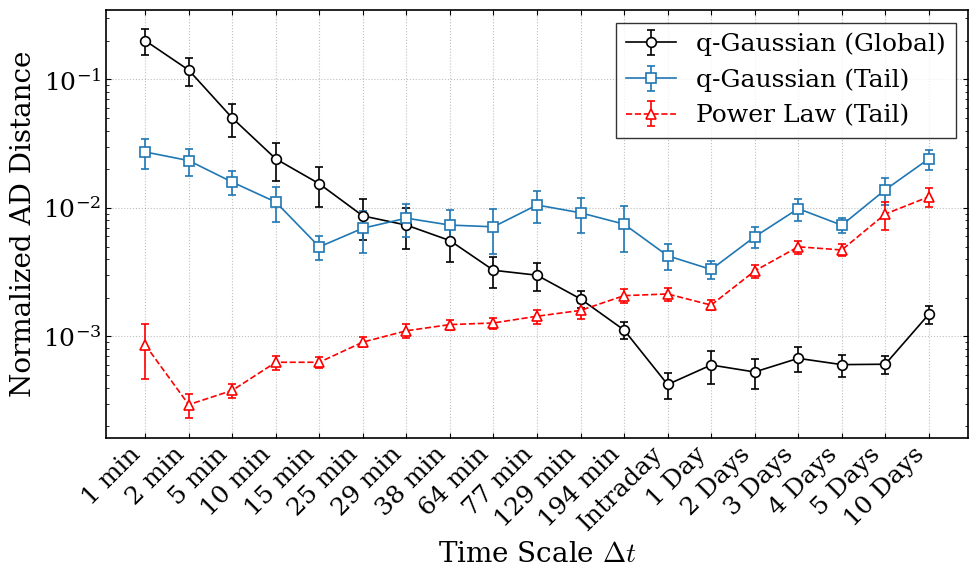}
        \caption{Behavior of the Normalized Anderson--Darling Statistic ($A^2/N$) -- U.S.}
        \label{fig:dir3}
    \end{subfigure}
    
    \caption{Statistics (Kolmogorov--Smirnov, Cram\'{e}r--von Mises, and
    Anderson--Darling) for different time scales $\Delta t$ (temporal lags of
    the log-returns) for U.S. and Brazilian stocks. The statistics are shown for
    three different fits: the $q$-Gaussian fitted to the entire dataset (global
    fit), the $q$-Gaussian fitted to the tail, and the power-law fit to the
    tail.}
    \label{fig:comparacao_6_imagens}
\end{figure}

Although a superficial reading of these statistics might suggest a
definitive victory for the power law in the extreme quantiles, it is
essential to ask whether this localized advantage is statistically
significant. The power law benefits from an in-sample optimization bias
because its parameters are estimated by MLE explicitly to minimize
discrepancies in the truncated tail, the q-Gaussian is constrained
by the body of the distribution.

\begin{figure}[tbph]
\centering
\par
\begin{subfigure}{0.48\textwidth}
        \centering
        \includegraphics[width=\linewidth]{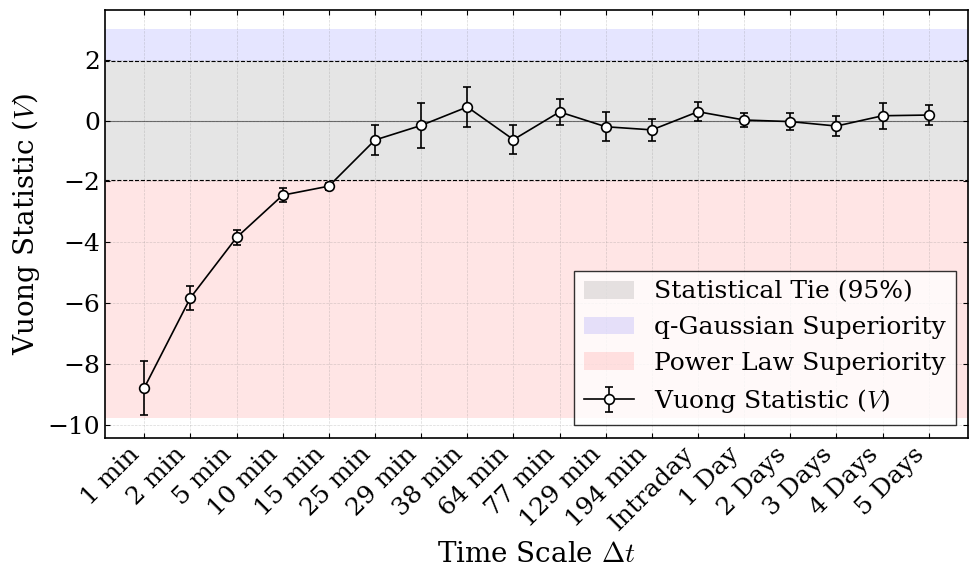}
        \caption{Behavior of the Vuong Statistic ($V$) for Brazilian Stocks.} 
        \label{fig:esq}
    \end{subfigure}\hfill 
\begin{subfigure}{0.48\textwidth}
        \centering
        \includegraphics[width=\linewidth]{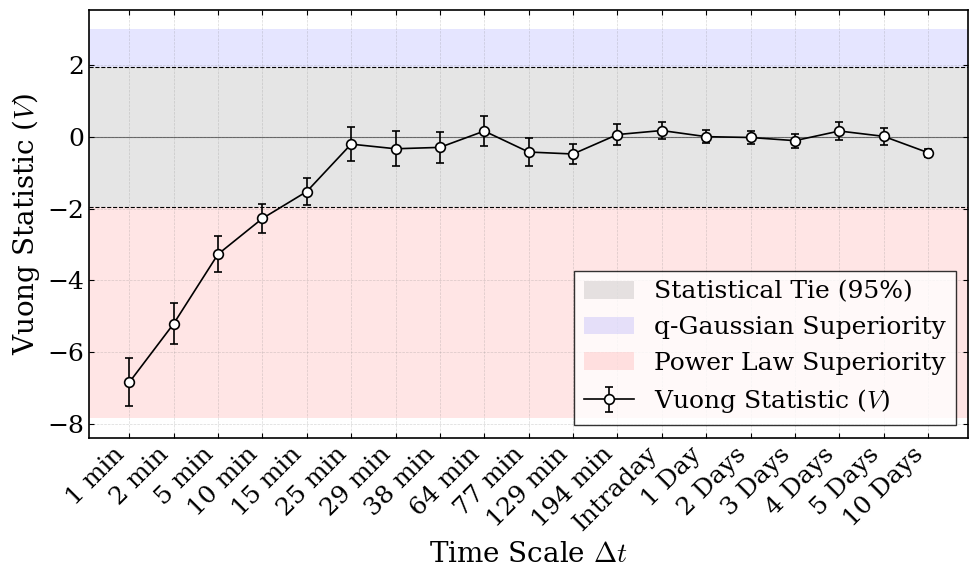}
        \caption{Behavior of the Vuong Statistic ($V$) for U.S. Stocks.}
        \label{fig:dir}
    \end{subfigure}
\caption{Pointwise comparison of the log-likelihood ratio using Vuong's $V$
statistic, considering only the tail region of the distribution. The shaded
central region denotes the non-rejection interval, within which the two tail
models are statistically indistinguishable.}
\label{fig:teste_vuong}
\end{figure}

To determine whether this apparent superiority is statistically significant,
we analyze the pointwise log-likelihood ratio using the Vuong test (see Fig. %
\ref{fig:teste_vuong}), considering only the tail region of the
distribution. The behavior of the statistic $V$ reveals a clear structural
transition. At ultra-high frequencies (for example, from 1 to 5 minutes),
the power-law model is clearly favored ($V<-1.96$). However, as the time
scale approaches approximately 15--25 minutes, $V$ rapidly converges toward
the non-rejection interval ($-1.96\leq V\leq 1.96$). This indicates that the
two tail models become statistically indistinguishable: the empirical data
no longer provide sufficient evidence to conclude that the power-law model
is superior to the $q$--Gaussian in describing the tail.

\begin{figure}[tbph]
\centering
\par
\begin{subfigure}{0.48\textwidth}
        \centering
        \includegraphics[width=\linewidth]{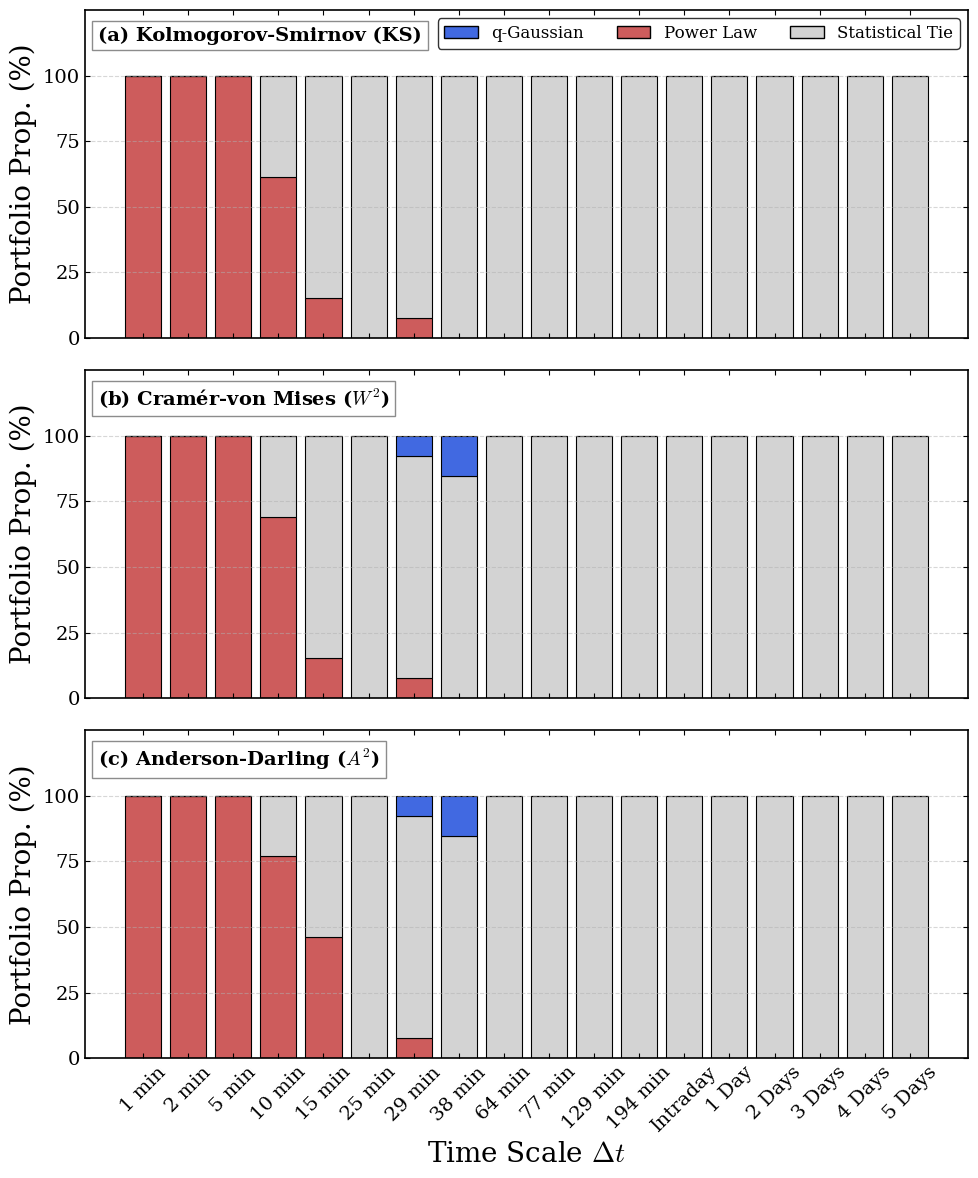}
        \caption{Bootstrap test for the Brazilian Stocks.}
        \label{fig:esq}
    \end{subfigure}\hfill 
\begin{subfigure}{0.48\textwidth}
        \centering
        \includegraphics[width=\linewidth]{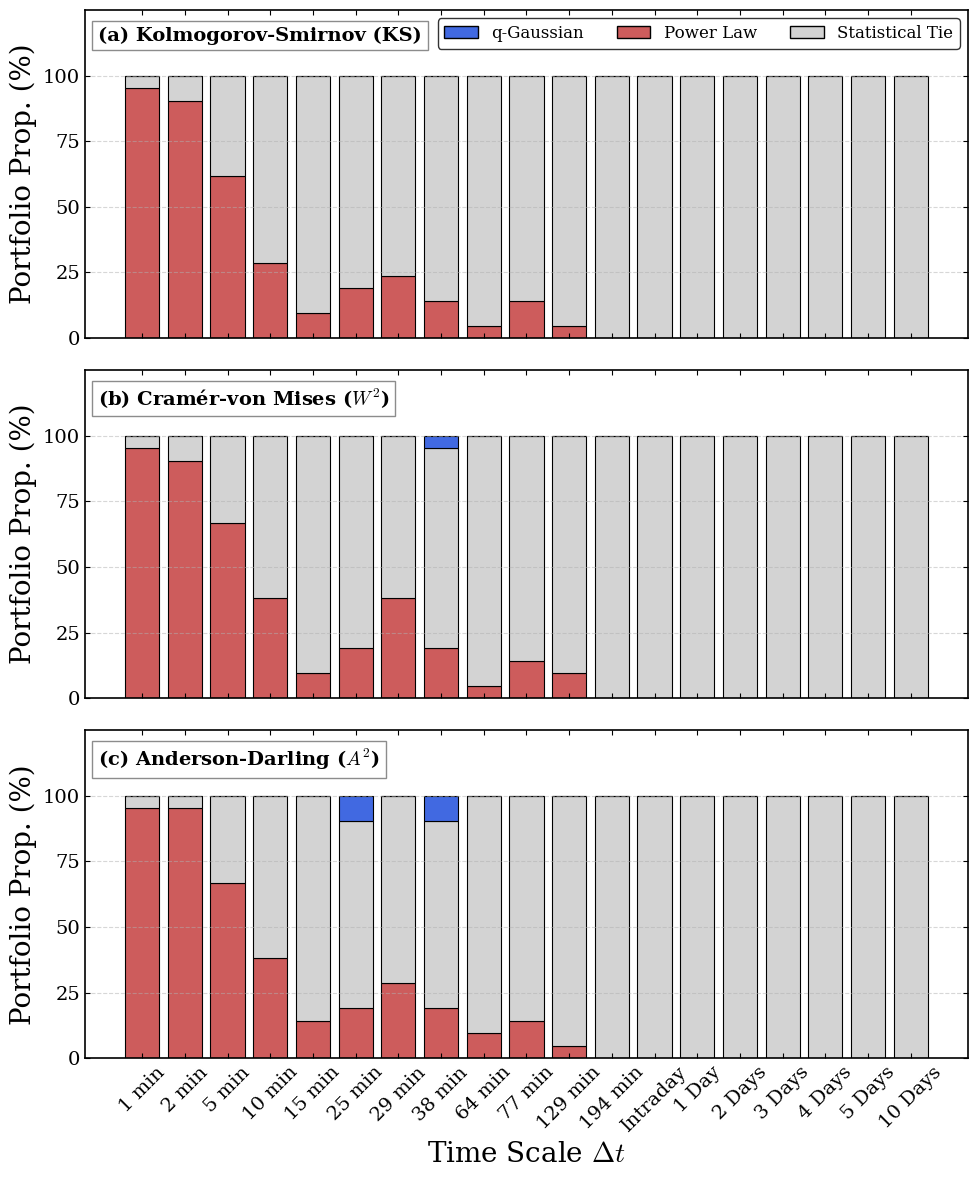}
        \caption{Bootstrap test for the U.S. Stock.}
        \label{fig:dir}
    \end{subfigure}
\caption{Bootstrap results for the three different statistics
Kolmogorov--Smirnov, Cram\'{e}r--von Mises, and Anderson--Darling, for both
Brazilian and U.S. stocks for the tail distribution. Red indicates cases in
which the power-law model is statistically favored, blue indicates cases in
which the $q$--Gaussian is favored, and gray denotes a statistical tie}
\label{fig:bootstrap}
\end{figure}

This loss of statistical distinguishability is further supported by
nonparametric bootstrap simulations ($2500$ iterations per asset), whose
results are shown in Fig. \ref{fig:bootstrap} for the three goodness-of-fit
statistics -- Kolmogorov--Smirnov, Cram\'{e}r--von Mises, and
Anderson--Darling -- for both Brazilian and U.S. stocks, considering only
the tail region of the distributions. In the figure, the color red indicates
cases in which the power-law model is statistically favored, blue indicates
cases in which the $q$--Gaussian is favored, and gray denotes a statistical
tie. The bootstrap results show that the dominance of the power law weakens
rapidly as the temporal aggregation scale increases. Starting at the
15-minute scale, the power-law model exhibits statistically significant
superiority for only a small fraction of the assets (approximately
20\%--30\%). As the aggregation scale increases further, the fraction of
assets for which the power law is significantly favored progressively
decreases and eventually vanishes. At the 194-minute scale, all assets
exhibit a statistical tie for all three goodness-of-fit statistics.

It is important to emphasize that the success of the $q$--Gaussian does not
arise merely by exclusion because statistical scarcity penalizes the power
law at macroscopic scales. The analytical quality of the model origined in
the noxtensive extension of the entropy is already clearly demonstrated in
the data-rich regime.

Within the mesoscale window (between 15 and 60 minutes), the number of
extreme observations in the tail is still of the order of thousands,
ensuring stability and high precision for the maximum-likelihood fits. Even
in this highly populated regime, the Vuong test (Fig.~\ref{fig:teste_vuong})
shows that the globally fitted $q$--Gaussian reaches a rigorous statistical
tie with the power law within the tail region.

This result is noteworthy from a modeling perspective. Whereas the power law
has the statistical advantage of being optimized exclusively for the tail,
discarding data below $x_{min}$, the q-Gaussian is subjected to a much
stricter constraint because it must simultaneously balance the probability
density in the body and the asymptotic decay. The fact that the $q$-Gaussian
matches the performance of the power law in the deep tail immediately after
the dissipation of microstructural noise ($\Delta t\geq 15$ min), and long
before data scarcity becomes a problem, shows that its mathematical
structure captures the behavior of financial returns as effectively as the
power law, while providing a substantially simpler and more practical fit.

\subsection{Considerations Regarding Cryptocurrency-Market Data}

\begin{figure}[htbp]
\centering
\par
\begin{subfigure}{0.48\textwidth}
        \centering
        \includegraphics[width=\linewidth]{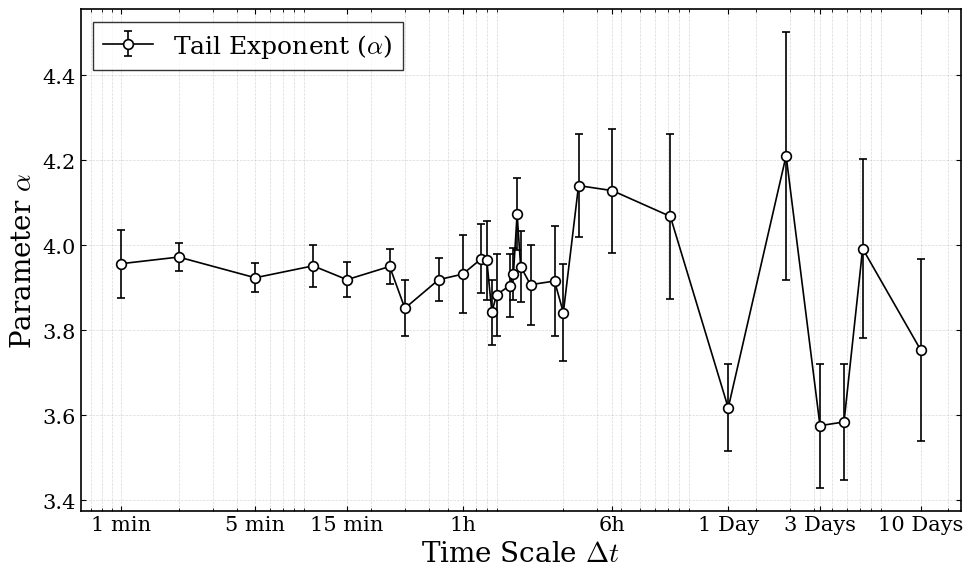}
        \caption{Behavior of the Power-Law Exponent ($\alpha$).}
        \label{fig:cripto_alfa}
    \end{subfigure}\hfill 
\begin{subfigure}{0.48\textwidth}
        \centering
        \includegraphics[width=\linewidth]{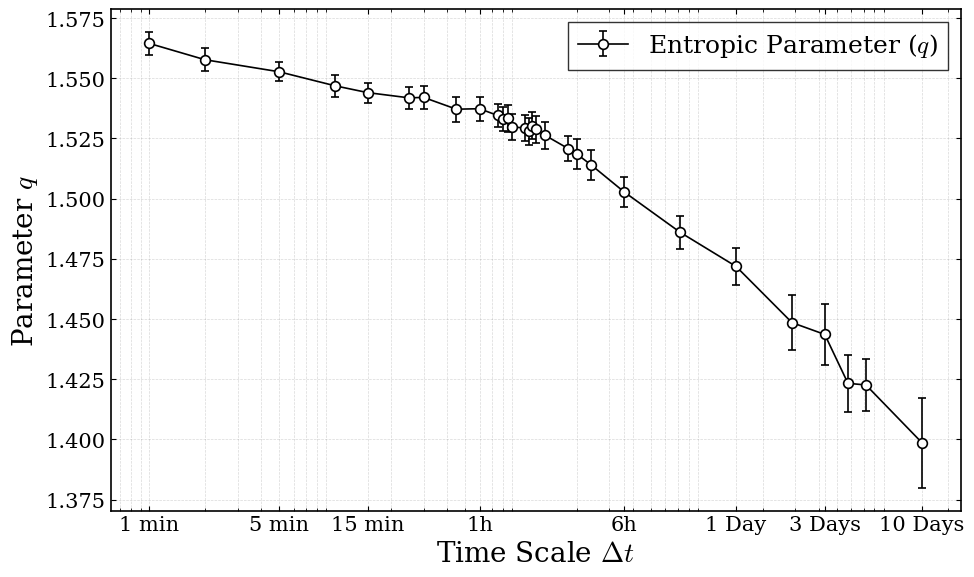}
        \caption{Behavior of the Entropic Parameter ($q$).}
        \label{fig:cripto_q}
    \end{subfigure}
\caption{Behavior of the descriptive parameters for the cryptocurrency
market as a function of the temporal aggregation window $\Delta t$. Error
bars represent the sample dispersion (standard deviation of the mean) across
the analyzed assets.}
\label{fig:criptos_alfa_q}
\end{figure}

Extending the analysis to the cryptocurrency ecosystem reveals a
phenomenological scenario that strongly corroborates the limitations of an
isolated power-law fit while also exposing unique mechanical features of
this market. Heavy-tailed return distributions and scale-dependent statistical
properties have likewise been documented for Bitcoin and other cryptocurrency
markets \cite{begusic2018,takaishi2018,drozdz2018}. As observed for the stock market, the behavior of $\alpha$
Fig.~\ref{fig:criptos_alfa_q}(a) fails to capture aggregational
Gaussianity. Instead of diverging to infinity ($\alpha\to\infty$), as
predicted by the Central Limit Theorem at macroscopic scales, the MLE
estimate fluctuates and stabilizes at finite values, predominantly between $%
3.6$ and $4.2$. This spurious stabilization again demonstrates that the
local-minimization criterion forces an algebraic-density fit over the
remaining marginal events, masking the actual transition of the system
toward normality.

By contrast, the entropic parameter $q$ of the q-Gaussian distribution [Fig.~%
\ref{fig:criptos_alfa_q}(b)] maps the relaxation dynamics of the
fluctuations continuously and coherently. Because the q-Gaussian fit uses
all the information available in the distribution, it is not affected by the
scarcity of extreme-tail data that penalizes the power law at broader time
windows. A crossover between two distinct linear regimes occurs at 
$\Delta t_c \approx 3 $h.

Moreover, the quantitative analysis of the nonextensivity parameter reveals
striking structural parallels and differences among the financial
ecosystems. The absolute values of $q$ for cryptocurrencies are very similar
to those observed in the U.S. stock market. This quantitative parity
indicates that, despite their operational differences, both markets exhibit
comparable levels of intermittency and frequencies of extreme events. On the
other hand, the magnitude of $q$ in cryptocurrencies is markedly higher than
in the Brazilian stock market. This discrepancy suggests that the
cryptoasset ecosystem shares the high complexity, informational volume, and
propensity for large shocks typical of the global U.S. market, while
differing from the less extreme behavior observed in the microstructure of
the Brazilian emerging market. A more detailed analysis may reveal
correlations between the U.S. and cryptocurrency markets.

\subsubsection{Absence of Structural Breaks: Continuous 24/7 Trading}

A fundamental mechanical difference emerges when the evolution of the
cryptocurrency parameter $q$ is compared with the previous stock-market
results. On traditional exchanges, a pronounced statistical anomaly was
observed at the one-day scale, generated by the \textit{overnight} effect:
information accumulated while the market was closed reintroduced volatility
into the system, producing a jump in the degree of nonextensivity.

For cryptocurrencies, however, this anomaly is entirely absent. Because this
ecosystem operates continuously, 24 hours a day and 7 days a week, it
imposes no institutional bottlenecks on information processing. As shown in
Fig.~\ref{fig:criptos_alfa_q}(b), the relaxation of $q$ follows a remarkably
smooth and continuous trajectory from intraday scales to multi-day scales.
The absence of structural breaks associated with closing and reopening hours
allows volatility to dissipate without interruption, resulting in a much
smoother and more consistent long-term Gaussianization process.

\subsection{Empirical Goodness of Fit for Cryptocurrencies}

\begin{figure}[htbp]
\centering
\par
\begin{subfigure}{0.48\textwidth}
        \centering
        \includegraphics[width=\linewidth]{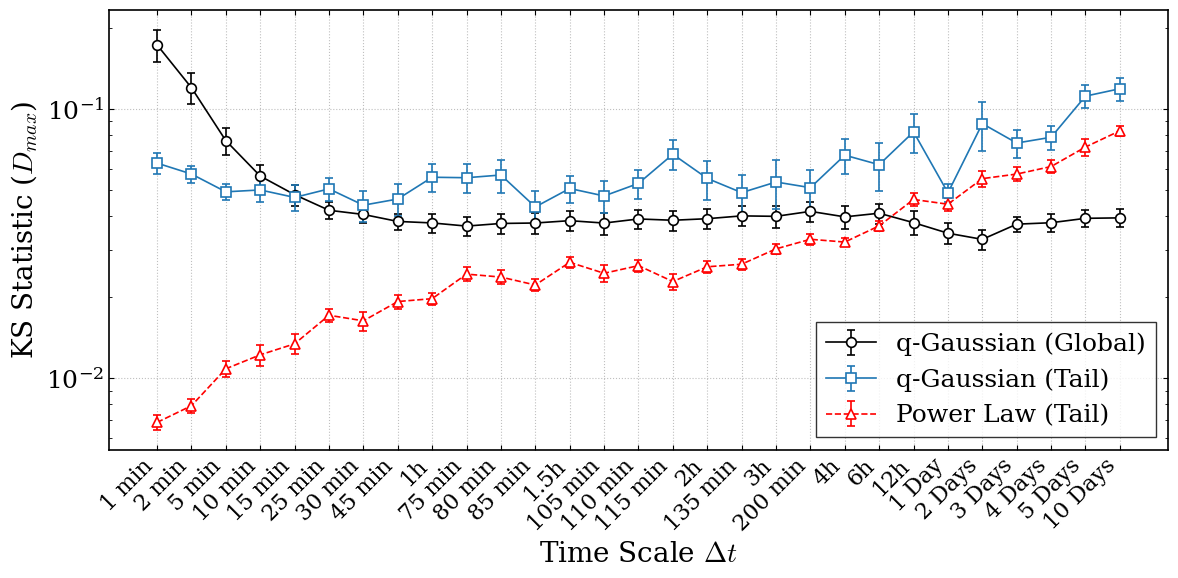}
        \caption{KS Distance ($D_{max}$).}
    \end{subfigure}\hfill 
\begin{subfigure}{0.48\textwidth}
        \centering
        \includegraphics[width=\linewidth]{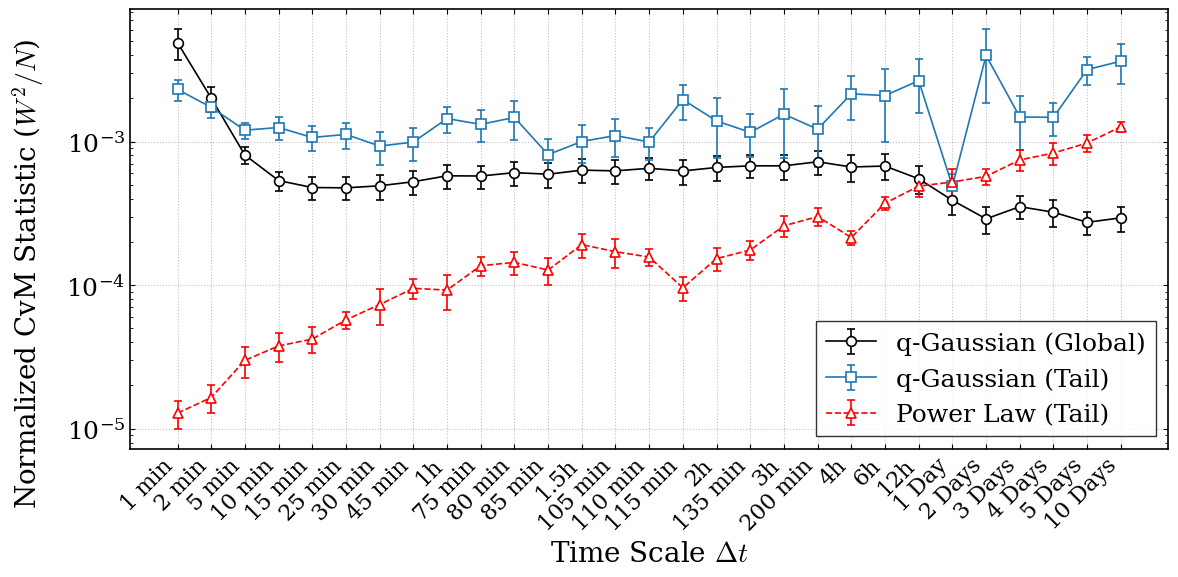}
        \caption{Normalized CvM Statistic ($W^2/N$).}
    \end{subfigure}
\par
\vspace{0.3cm}
\par
\begin{subfigure}{0.48\textwidth}
        \centering
        \includegraphics[width=\linewidth]{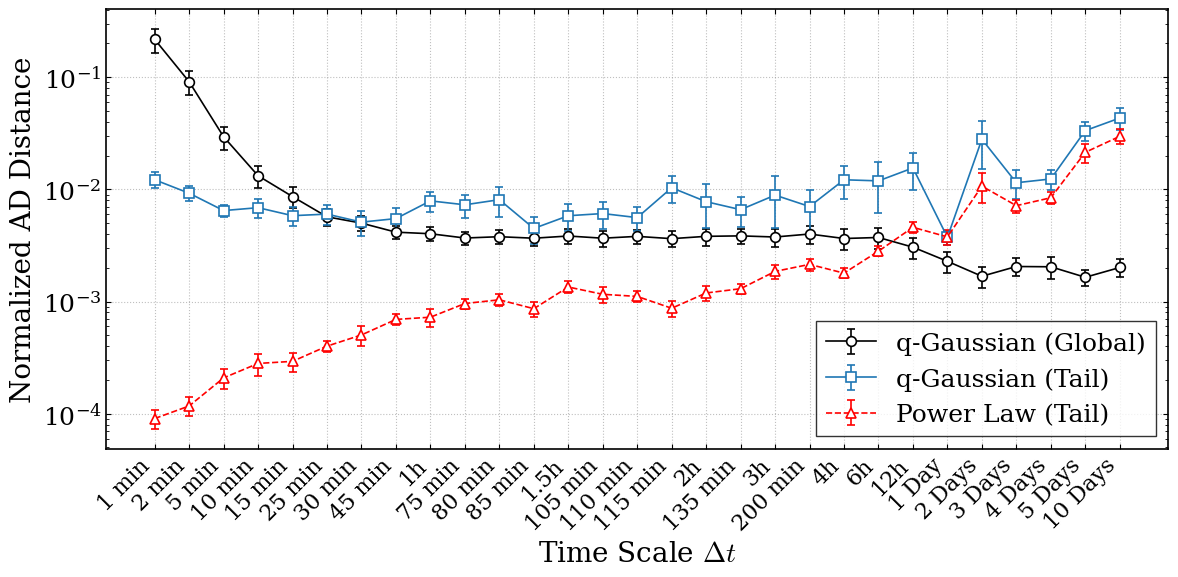}
        \caption{Normalized AD Statistic ($A^2/N$).}
    \end{subfigure}\hfill 
\begin{subfigure}{0.48\textwidth}
        \centering
        \includegraphics[width=\linewidth]{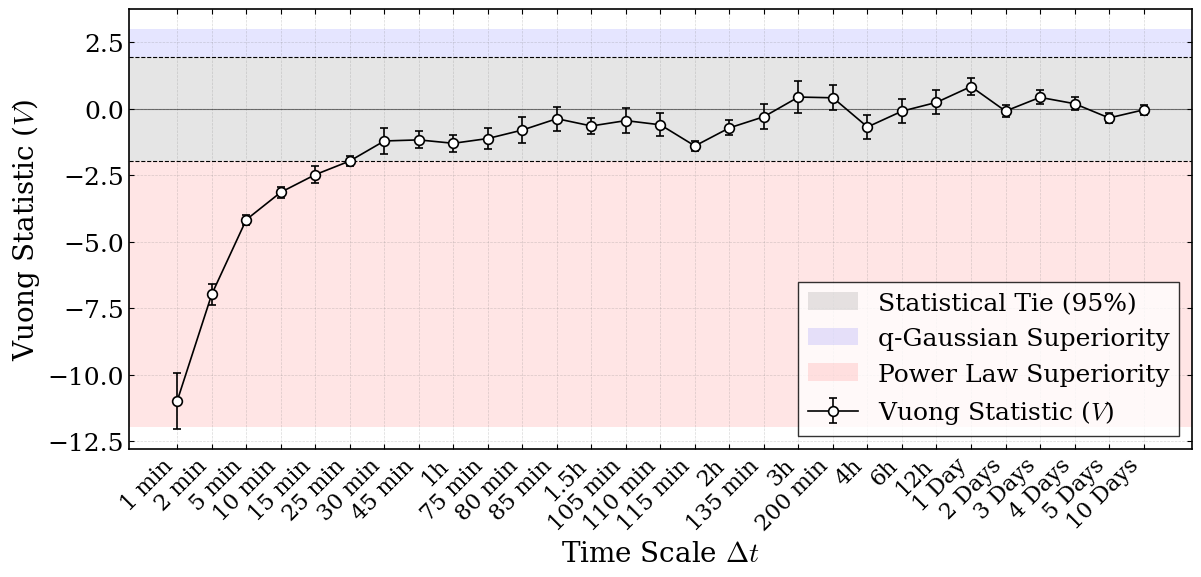}
        \caption{Vuong Statistic ($V$).}
        \label{fig:cripto_vuong}
    \end{subfigure}
\caption{Evolution of the goodness-of-fit metrics and log-likelihood ratio
(Vuong test) for the cryptocurrency ecosystem across multiple temporal
resolutions.}
\label{fig:criptos_gof}
\end{figure}

The evolution of the goodness-of-fit metrics for cryptocurrencies reveals a
convergence pattern analogous to that found for stocks. Analysis of the
geometric Kolmogorov--Smirnov (KS), Cramér--von Mises (CvM),
and Anderson--Darling (AD) distances, shown in Fig.~\ref{fig:criptos_gof},
indicates that the global q-Gaussian fit improves continuously as
high-frequency microstructural noise dissipates.

At the same time, the absolute error of the power law in the extreme region
deteriorates monotonically as $\Delta t$ increases. Nevertheless, as in the
traditional market, the power-law error remains nominally smaller than that
of the q-Gaussian when evaluated strictly in the tail.

\begin{figure}[htbp]
\centering
\includegraphics[width=\textwidth]{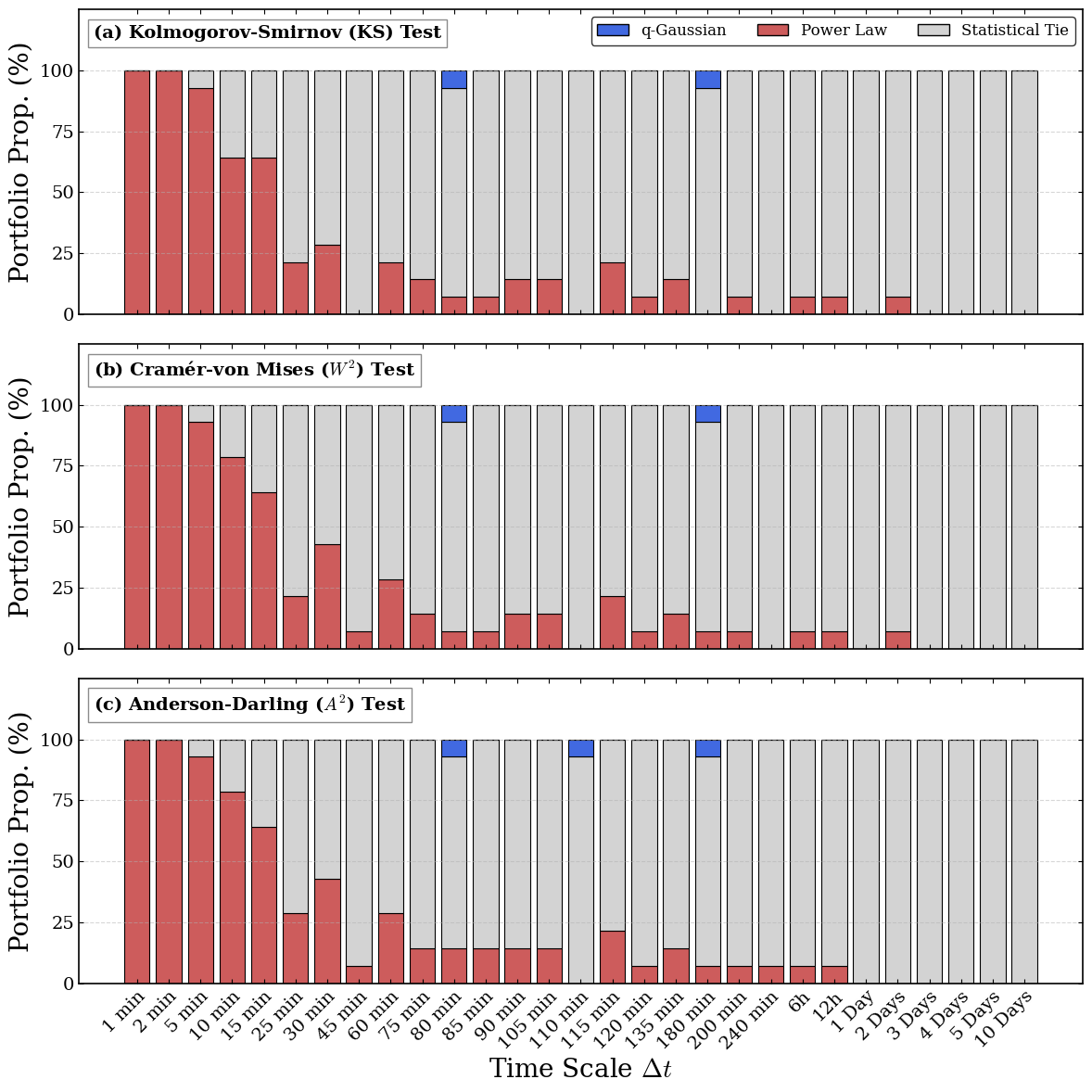}
\caption{Nonparametric \textit{bootstrap} validation ($2500$ iterations) of
geometric goodness of fit in the tail regime for the cryptocurrency market.
The bar charts show the proportion of the asset portfolio for which the
globally fitted q-Gaussian (blue), the locally fitted power law (red), or a
statistical tie (gray) was superior according to the KS, CvM, and AD
statistics as a function of the time scale $\Delta t$.}
\label{fig:criptos_bootstrap}
\end{figure}

The illusion of this apparent geometric superiority of the power law is
rigorously dismantled by applying the Vuong closeness test (Fig.~\ref%
{fig:cripto_vuong}) and, more definitively, by the nonparametric \textit{%
bootstrap} simulations ($2500$ iterations), whose results are summarized in
Fig.~\ref{fig:criptos_bootstrap}.

The observed structural dynamics are clear: at very short time scales ($%
\Delta t$ from 1 to 2 minutes), where microstructural effects assign
disproportionate weight to the core of the distribution, the power law---by
discarding this core and restricting the domain to $x\ge x_{min}$%
---unanimously dominates the tail fit for 100\% of the portfolio. This
dominance, however, breaks down rapidly. Beginning with the 5-minute window,
the proportion of strict power-law victories falls sharply, and the system
continuously transitions toward a statistical-tie regime, indicating that
sampling uncertainty becomes greater than the difference in error between
the models.

At mesoscales between 25 and 60 minutes, statistical ties already account
for the overwhelming majority of cases for all three geometric metrics (KS,
CvM, and AD). As temporal aggregation proceeds, the power-law advantage
disappears entirely. From the 1440-minute (one-day) scale onward, an
absolute statistical tie is established for 100\% of the analyzed assets.

Despite being subject to the stringent mathematical requirement of
simultaneously describing the central dynamics and the asymptotic decay, the
q-Gaussian reaches and maintains statistically indistinguishable performance
in representing extreme events once the initial microstructural noise has
been overcome. This \textit{bootstrap} validation shows that the nominal
error advantage of the power law observed at macroscopic scales does not
reflect an underlying strictly fractal mechanism; rather, it is
fundamentally an artifact generated by the intrinsic noise of a finite
sample in the deep tail.

\begin{figure}[htbp]
\centering
\par
\begin{subfigure}{0.32\textwidth}
        \centering
        \includegraphics[width=\linewidth]{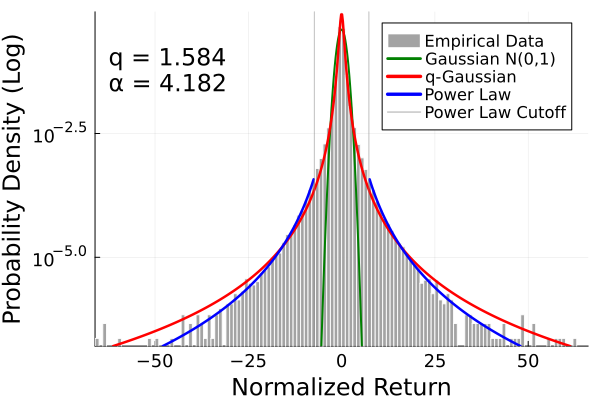}
        \caption{1 minute.}
        \label{fig:btc_1m}
    \end{subfigure}\hfill 
\begin{subfigure}{0.32\textwidth}
        \centering
        \includegraphics[width=\linewidth]{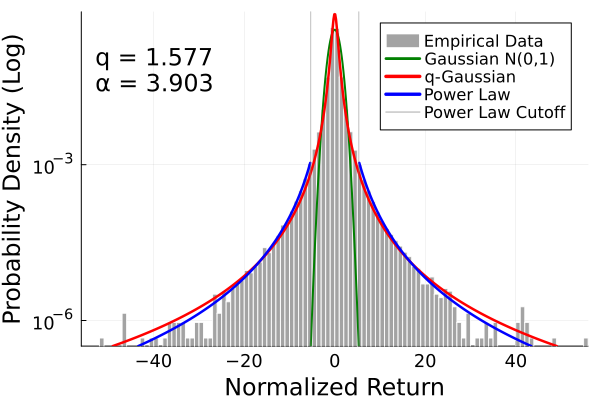}
        \caption{2 minutes.}
        \label{fig:btc_2m}
    \end{subfigure}\hfill 
\begin{subfigure}{0.32\textwidth}
        \centering
        \includegraphics[width=\linewidth]{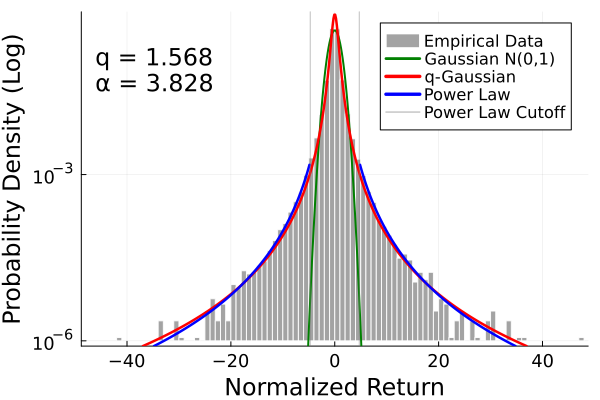}
        \caption{5 minutes.}
        \label{fig:btc_5m}
    \end{subfigure}
\par
\vspace{0.4cm} 
\par
\begin{subfigure}{0.32\textwidth}
        \centering
        \includegraphics[width=\linewidth]{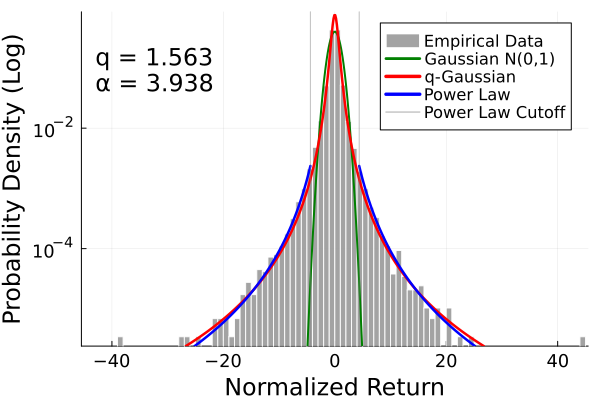}
        \caption{15 minutes.}
        \label{fig:btc_15m}
    \end{subfigure}\hfill 
\begin{subfigure}{0.32\textwidth}
        \centering
        \includegraphics[width=\linewidth]{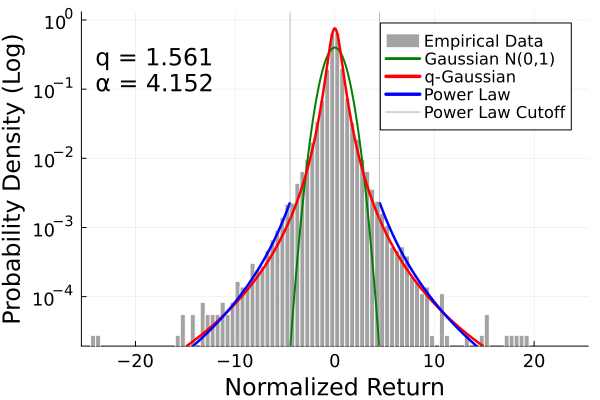}
        \caption{60 min (1 hour).}
        \label{fig:btc_60m}
    \end{subfigure}\hfill 
\begin{subfigure}{0.32\textwidth}
        \centering
        \includegraphics[width=\linewidth]{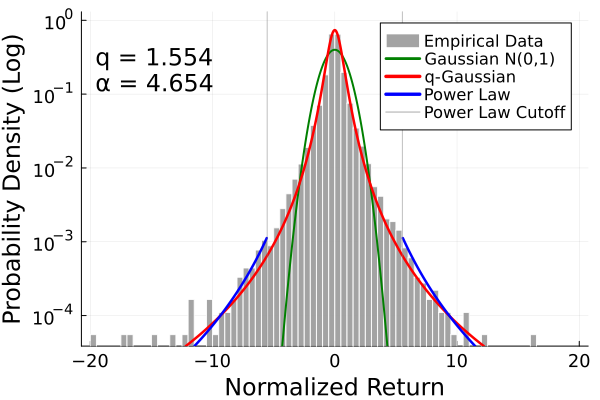}
        \caption{120 min (2 hours).}
        \label{fig:btc_120m}
    \end{subfigure}
\par
\vspace{0.4cm} 
\par
\begin{subfigure}{0.32\textwidth}
        \centering
        \includegraphics[width=\linewidth]{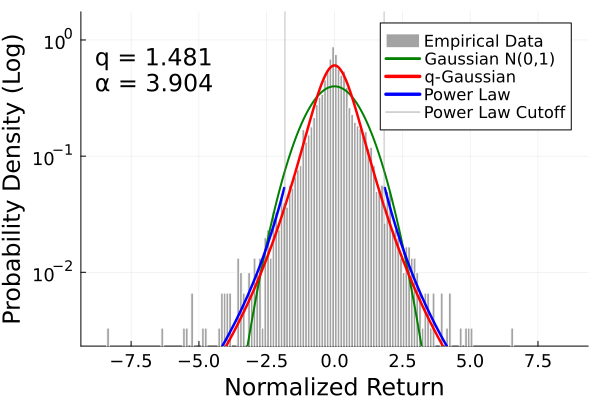}
        \caption{1.440 min (1 day).}
        \label{fig:btc_1440m}
    \end{subfigure}\hfill 
\begin{subfigure}{0.32\textwidth}
        \centering
        \includegraphics[width=\linewidth]{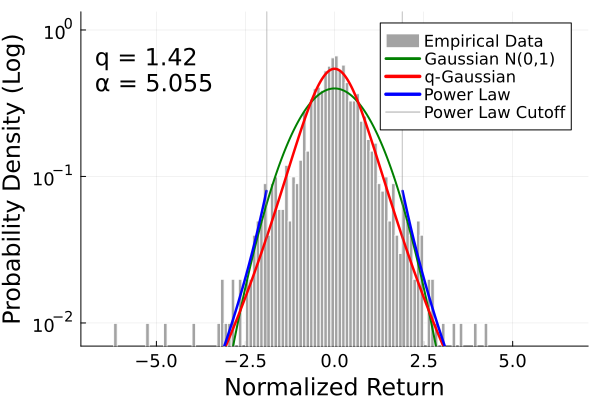}
        \caption{4.320 min (3 days).}
        \label{fig:btc_4320m}
    \end{subfigure}\hfill 
\begin{subfigure}{0.32\textwidth}
        \centering
        \includegraphics[width=\linewidth]{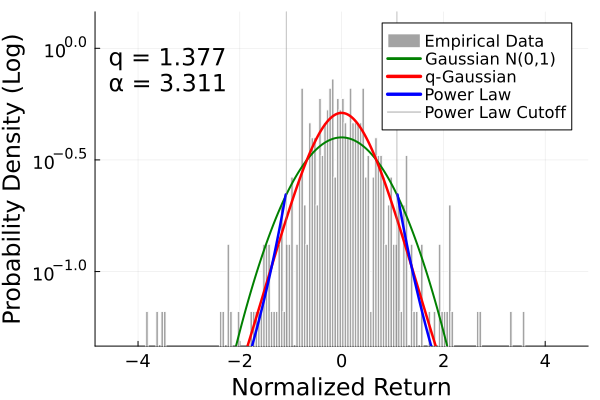}
        \caption{14.400 min (10 days).}
        \label{fig:btc_14400m}
    \end{subfigure}
\caption{Evolution of the Bitcoin (BTC) time series across different time
windows, ranging from 1 minute to 10 days.}
\label{fig:grid_btc}
\end{figure}

\subsection{Computational Complexity of the Fitting Algorithms}

When dealing with high-frequency financial time series, computational
efficiency becomes as critical as statistical robustness. The volume of
intraday data requires algorithms that scale efficiently. To assess
rigorously the performance of the proposed methods, we analyze their
asymptotic computational complexity using Big-O notation, where $N$ denotes
the total number of observations in the empirical dataset.

\subsubsection{Complexity of Power-Law Fitting (KS Minimization and MLE)}

The standard procedure for fitting a power-law distribution involves
determining the optimal lower bound $x_{min}$ and subsequently estimating
the scaling parameter $\alpha$. The optimal cutoff $x_{min}^*$ is the value
that minimizes $D$, thereby ensuring that the empirical tail follows the
theoretical power-law model as closely as possible.

Algorithmically, this requires iterating over every unique value in the
dataset and treating each one as a candidate for $x_{min}$. The
computational steps are as follows:

\begin{enumerate}
\item \textbf{Sorting:} The data must be sorted to calculate the empirical
cumulative distribution function (CDF), which has a time complexity of $%
O(N\log N)$.

\item \textbf{Iterative Evaluation:} For each candidate $x_{min}$, the
algorithm calculates the maximum-likelihood estimate (MLE) of $\alpha$ and
evaluates the Kolmogorov--Smirnov (KS) distance over the remaining tail. If
the candidate tail contains $n$ points, calculating the distance requires $%
O(n)$ operations.
\end{enumerate}

In the worst-case scenario (continuous data with no repeated values), the
algorithm evaluates tails of sizes $N,N-1,N-2,\dots,1$. The total number of
operations in this stage scales as an arithmetic progression and therefore, 
performing the basic power-law fit once requires $O(N^2)$ operations.

\subsubsection{Complexity of the Moment-Ratio Method (q-Gaussian)}

By contrast, the moment-ratio method used to fit the nonextensive q-Gaussian
distribution decouples the optimization procedure from the time-series
length $N$. The algorithm operates in two distinct stages:

\begin{enumerate}
\item \textbf{Empirical Evaluation:} The algorithm calculates the absolute
moments $M_k=\langle|x|^k\rangle$ for a predefined fractional vector of
length $K$. Scanning a dataset of size $N$ to calculate a single moment
costs $O(N)$. Therefore, extracting all empirical moments requires $O(K\cdot
N)$ operations.

\item \textbf{Theoretical Optimization:} The nonlinear least-squares
optimization iteratively proposes values of the entropic index $q$. For each
proposed $q$, the theoretical moment ratio $R_q(k)$ is calculated over the
vector of length $K$ using algebraic Gamma functions, at a cost of $O(K)$.
If the optimizer requires $I$ iterations to converge, the cost of this stage
is $O(I\cdot K)$.
\end{enumerate}

The total computational complexity of the moment-ratio method is the sum of
the two stages: $O(K \cdot N) + O(I \cdot K)$. Because the number of fractional 
moments ($K$) and the number of optimization iterations ($I$) are small fixed 
constants that do not grow with dataset size, they may be omitted asymptotically. 
The final complexity of fitting the q-Gaussian distribution therefore reduces 
strictly to $O(N)$.

\subsubsection{Algorithmic Verdict}

The transition from KS-based power-law fitting to the moment-ratio method
represents a change from quadratic computational complexity, $O(N^2)$, to
linear complexity, $O(N)$. Whereas the traditional approach is fundamentally
limited by the bottleneck of repeatedly reevaluating the empirical CDF over
subsets of the data, the q-Gaussian methodology evaluates global statistical
properties only once. This linear scaling makes the q-Gaussian approach not
only statistically robust for mapping heavy tails without rigid cutoff
points, but also substantially more suitable for large-scale econophysics
applications.

\section{Discussion, Conclusions, and Final Remarks}

\label{Sec:Conclusions}

The detailed investigation of financial-return dynamics across multiple time
scales ($\Delta t$) reveals that traditional heavy-tail modeling, often
based on pure power laws, faces severe analytical and interpretative
limitations when confronted with aggregational Gaussianity. Throughout this
study, we have shown that the standard tail-estimation protocol---based on
maximum-likelihood estimation (MLE) coupled with local minimization of the
Kolmogorov--Smirnov (KS) distance---creates a critical methodological
artifact. Because the algorithm actively seeks to isolate the extreme subset
($x\ge x_{min}$) that best matches an algebraic density, it imposes a
spurious stabilization of the exponent $\alpha$ at finite values.
Consequently, the power law becomes ``blind'' to the macroscopic dynamics of
the system, failing to capture the convergence imposed by the Central Limit
Theorem and incorrectly suggesting the survival of strict scale-invariance
properties in temporal regimes in which the market is already undergoing a
clear normalization process.

By contrast, nonextensive statistical mechanics, implemented through the
q-Gaussian distribution, emerges as a substantially more robust and reliable
analytical framework for high-frequency financial modeling. By adopting a
global parameterization based on the fractional moment-ratio method, the
$q$-Gaussian eliminates the need for dynamic cutoff thresholds ($x_{min}$) and
the arbitrary discarding of data. The Tsallis index $q$ acts as a unified
morphological descriptor, accurately mapping the continuous relaxation of
anomalous fluctuations toward the Gaussian boundary ($q\to1$). The ability
to use the information density in the body of the distribution to constrain
the decay of its tails strongly mitigates the finite-size effects that
undermine the reliability of pure-tail models at large aggregation windows.

The apparent geometric superiority of the power law in the tail region was
rigorously dismantled through the Vuong closeness test and nonparametric 
\textit{bootstrap} simulations. The nominal advantage of the power law was
found to be restricted almost exclusively to the very-short-term regime (the
first few minutes), where microstructural noise---including frictions and
tick size---dominates the core of the distribution. Once this high-frequency
inertia dissipates, the q-Gaussian reaches and maintains an absolute
statistical tie in the representation of extreme events. The bootstrap
results definitively show that, at mesoscopic and macroscopic scales,
natural sampling variation exceeds any descriptive difference between the
models, demonstrating that the structural flexibility of the q-Gaussian
accommodates extreme volatility with accuracy comparable to that of a
dedicated tail model despite being subject to much stricter mathematical
constraints.

By extending the comparative analysis from traditional stock markets to the
cryptocurrency ecosystem, we reveal how the trading architecture itself
determines the rate of entropic dissipation. On stock exchanges, daily
trading interruptions generate structural breaks marked by the \textit{%
overnight effect}. The accumulation of latent informational shocks abruptly
reinjects volatility into the system at each reopening, appearing as
anomalous jumps in $q$ and interrupting intraday Gaussianization. In sharp
contrast, the continuous and uninterrupted flow of cryptocurrency trading
frees this market from such institutional bottlenecks. The absence of
closing periods allows volatility to flow without restriction, producing a
relaxation trajectory toward statistical equilibrium that is markedly
smoother and more consistent over multiple days.

Beyond the temporal relaxation dynamics, the absolute magnitude of $q$
proved to be an intrinsic signature of the complexity of each financial
ecosystem. Cryptocurrencies were found to operate at levels of
nonextensivity quantitatively similar to those of the U.S. stock market,
indicating comparable levels of intermittency and propensity for extreme
shocks. Both, however, exhibit values of $q$ significantly higher than those
found in the Brazilian market. This stratification suggests that, although
the emerging market also preserves anomalous behavior and heavy tails, its
microstructure generates extreme fluctuations of smaller magnitude and
intensity than those associated with the volume, liquidity, and information
density of the global U.S. markets and the decentralized cryptoasset network.

Finally, the empirical verdict is reinforced by an overwhelming
computational advantage. The transition from statistical inference based on
empirical CDF minimization (power law) to the mathematical evaluation of
global moments (q-Gaussian) improves algorithmic efficiency from quadratic
complexity, $O(N^2)$, to linear complexity, $O(N)$. In contemporary
econophysics, where financial databases easily exceed millions of
tick-by-tick observations, this asymptotic reduction is essential.

In summary, the q-Gaussian not only provides a more phenomenologically
accurate account of the stylized facts of complex markets, capturing both
central equilibrium behavior and tail criticality, but does so with strict
statistical validation and optimal computational scalability. Monitoring the
parameter $q$ therefore emerges as a promising and highly informative
alternative for detecting regime changes, shock absorption, and the
characterization of financial systems.

\begin{figure}[htbp]
    \centering
    \begin{subfigure}[t]{0.48\textwidth}
        \centering
        \includegraphics[width=\linewidth]{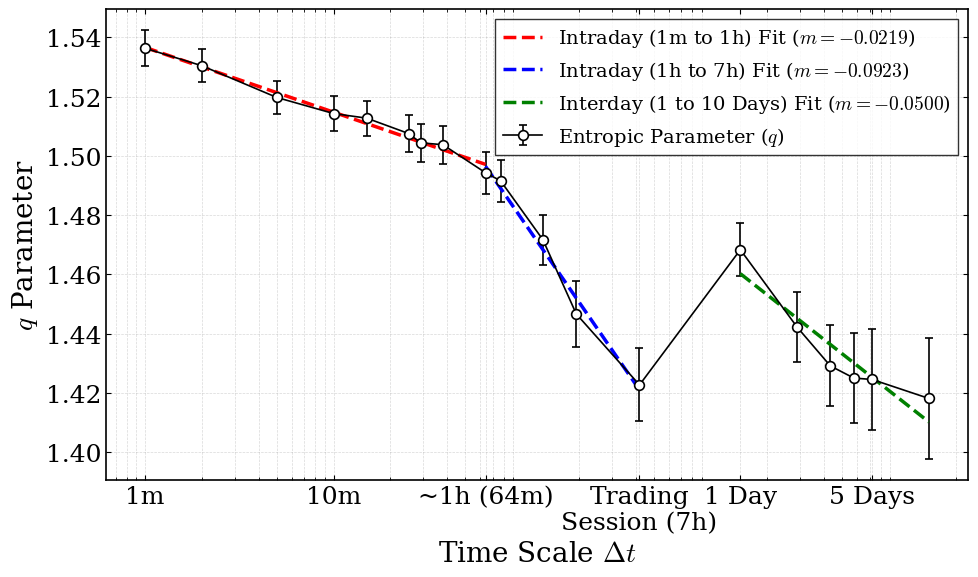}
        \caption{$q$-evolution regimes in the U.S. stock market.}
        \label{fig:regimes_eua}
    \end{subfigure}\hfill
    \begin{subfigure}[t]{0.48\textwidth}
        \centering
        \includegraphics[width=\linewidth]{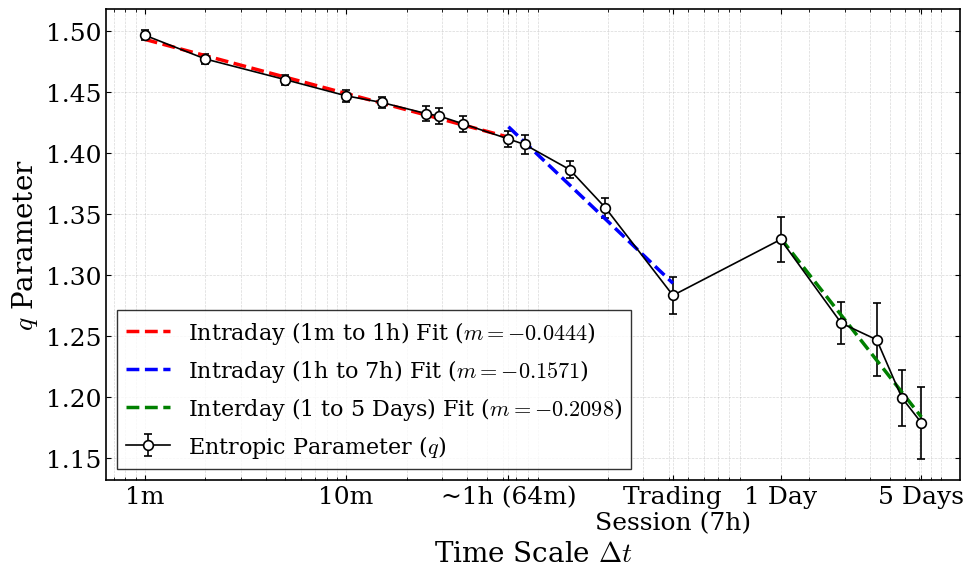}
        \caption{$q$-evolution regimes in the Brazilian stock market.}
        \label{fig:regimes_br}
    \end{subfigure}

    \vspace{0.4cm} 

    \begin{subfigure}[t]{0.48\textwidth}
        \centering
        \includegraphics[width=\linewidth]{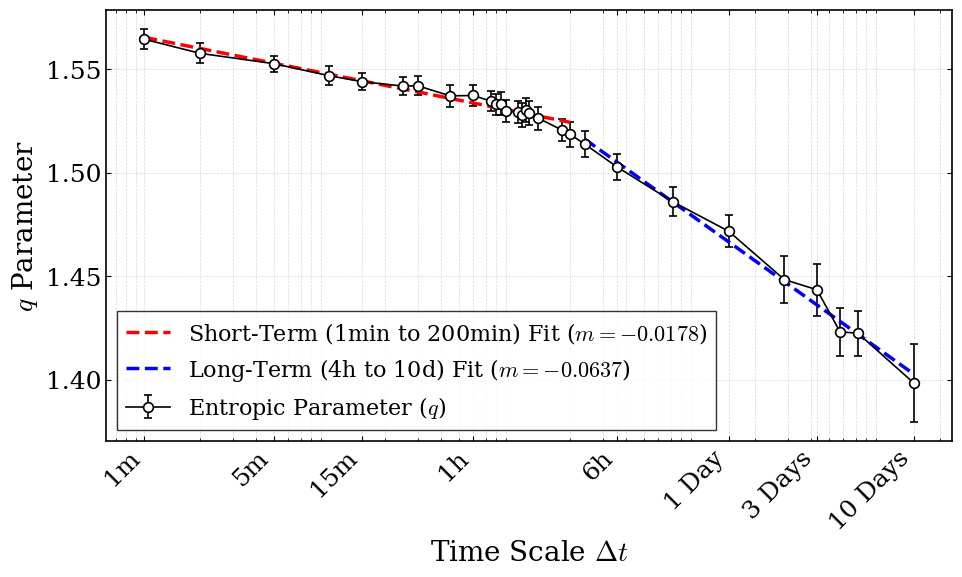}
        \caption{$q$-evolution regimes in the cryptocurrencies market.}
        \label{fig:regimes_criptos}
    \end{subfigure}

    \caption{Behavior of the Tsallis entropic parameter ($q$) and its corresponding qualitative local linear fits ($q = m \log \Delta t + c$) in the semi-log plot for the Brazilian, Cryptocurrency, and U.S. markets, highlighting distinct regime crossovers across intraday and interday scales.}
    \label{fig:tres_imagens_layout}
\end{figure}

Additionally, we observed a change in the $q$ evolution rate over time, as is shown in Fig \ref{fig:tres_imagens_layout}. To highlight the different regimes, we made a qualitative linear fitting in the semi-log plot and realize a pattern of dependence of the $q$ parameter on the log of time. Although a power-law fit also performs well due to its mathematical similarity for values close to $1$, the log-linear approach clearly demonstrates this change of regime.

For Brazilian and American stock market returns (Fig. \ref{fig:regimes_br} and Fig \ref{fig:regimes_eua}), until 1 hour both sustain a slower decrease compared to the following time intervals. Both markets present a steeper decrease of $q$ after 1 hour, with the Brazilian market showing a much more aggressive intraday decay ($m = -0.1571$) than the US market ($m = -0.0923$). Even though the overnight break blurs the perception, in both markets it is clear that this change of regime persists.

In the cryptocurrency evolution (Fig. \ref{fig:regimes_criptos}), this change is even clearer. Since there is no overnight effect here, once the system changes, it consistently sustains the new regime. Unlike the stock market, the break occurs approximately at 3 hours and 30 minutes, meaning that the initial slow decay is sustained longer than in equity markets. The decay rate shifts from $m = -0.0178$ to $m = -0.0637$, demonstrating a clear regime change across aggregation scales.

Regarding the approach to the Gaussian regime ($q = 1$), all three markets start with similar parameters at the 1-minute scale ($q \approx 1.50$ to $1.56$). However, the Brazilian market transitions to macroscopic behavior much faster, dropping to $q \approx 1.18$ in 5 days. In contrast, the US and cryptocurrency markets remain around $q \approx 1.40$ over similar macroscopic periods.

\begin{figure}[htbp]
    \centering
    
    \begin{subfigure}{0.48\textwidth}
        \centering
        \includegraphics[width=\linewidth]{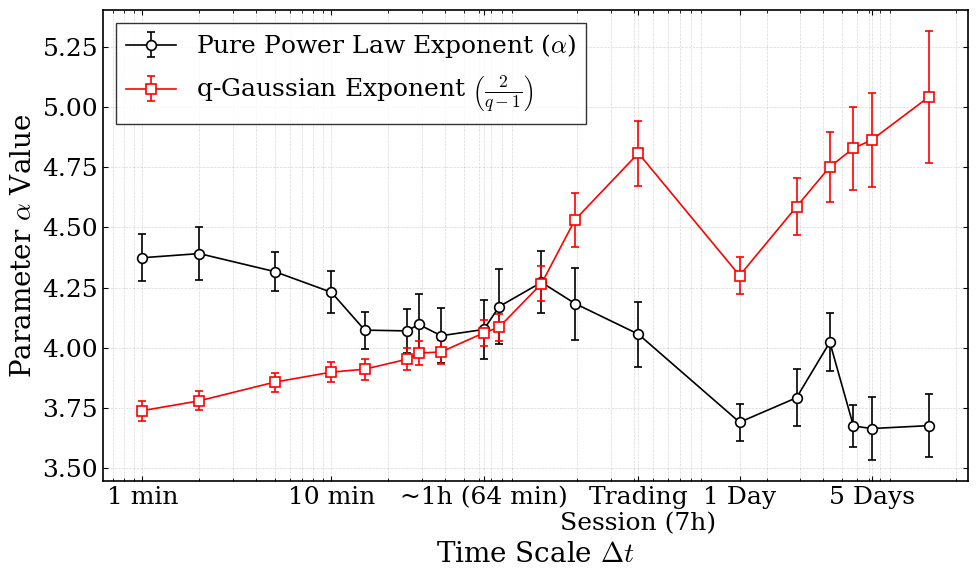}
        \caption{Individual evolution of each parameter $\alpha_q$ and $\alpha_{PL}$ for U.S. stocks.}
        \label{fig:comp_eua}
    \end{subfigure}
    \hfill
    \begin{subfigure}{0.48\textwidth}
        \centering
        \includegraphics[width=\linewidth]{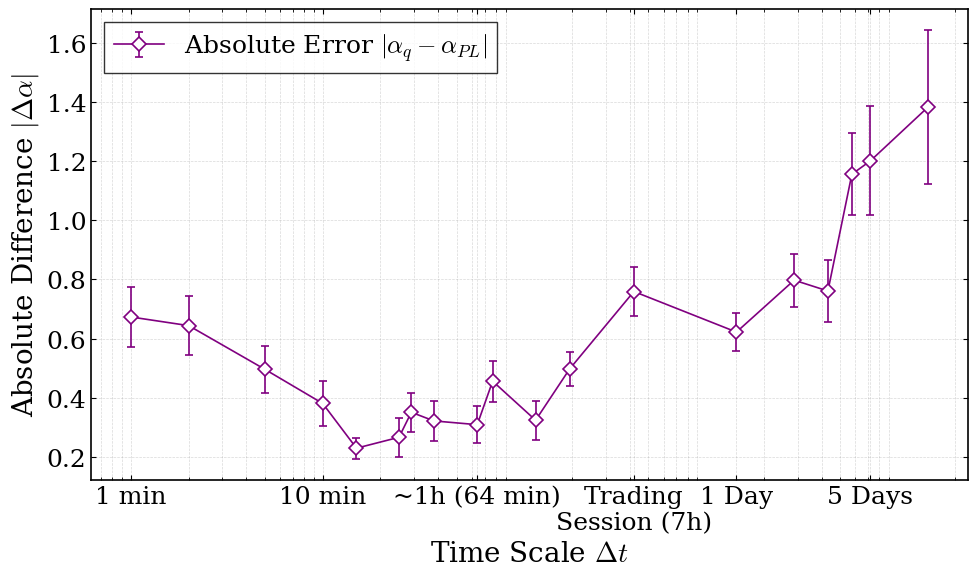}
        \caption{Absolute difference $|\alpha_q - \alpha_{PL}|$ for U.S. stock market.}
        \label{fig:dif_eua}
    \end{subfigure}
    
    \vspace{0.5cm} 
    
    \begin{subfigure}{0.48\textwidth}
        \centering
        \includegraphics[width=\linewidth]{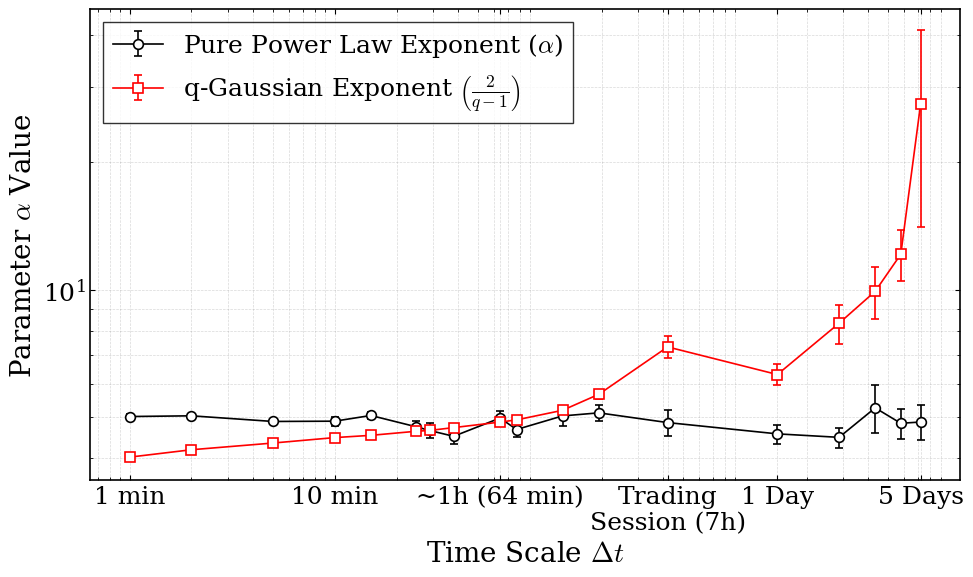}
        \caption{Individual evolution of each parameter $\alpha_q$ and $\alpha_{PL}$ for Brazilian stocks.}
        \label{fig:comp_brasil}
    \end{subfigure}
    \hfill
    \begin{subfigure}{0.48\textwidth}
        \centering
        \includegraphics[width=\linewidth]{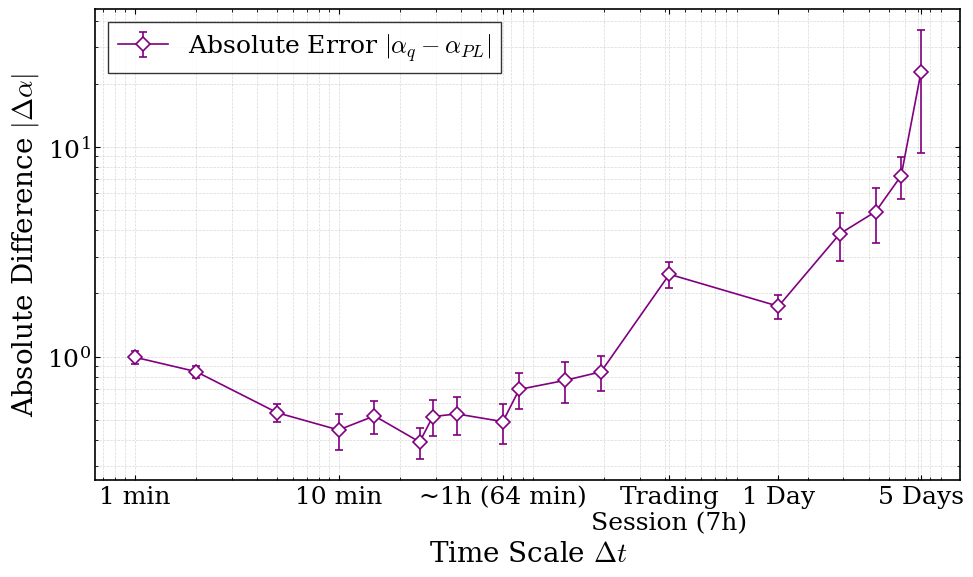}
        \caption{Absolute difference $|\alpha_q - \alpha_{PL}|$ for Brazilian stock market.}
        \label{fig:dif_brasil}
    \end{subfigure}
    
    \vspace{0.5cm} 
    
    \begin{subfigure}{0.48\textwidth}
        \centering
        \includegraphics[width=\linewidth]{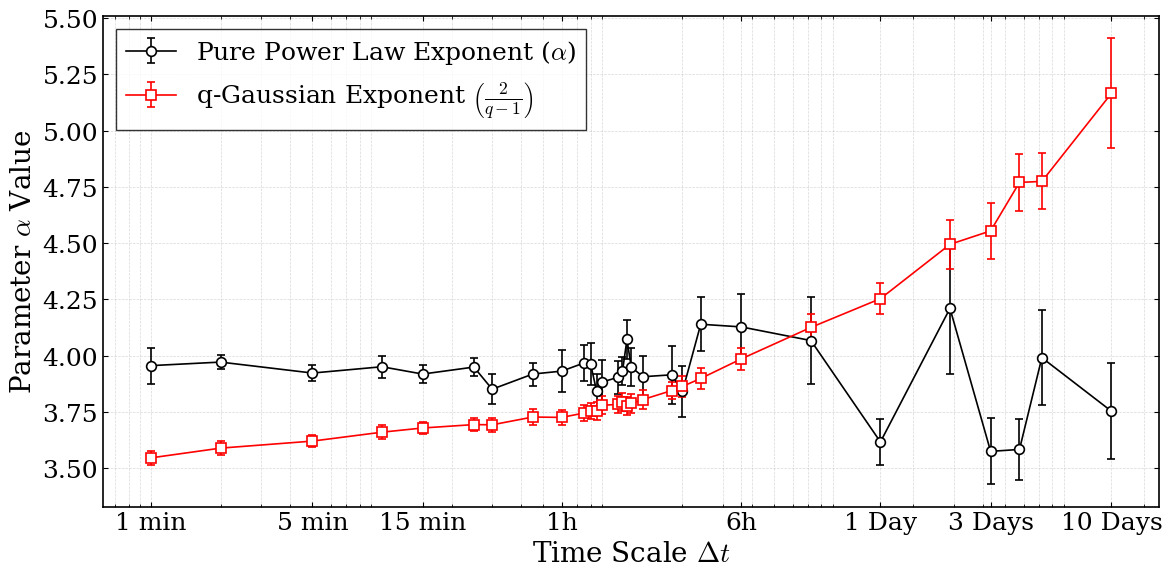}
        \caption{Individual evolution of each parameter $\alpha_q$ and $\alpha_{PL}$ for cryptocurrencies.}
        \label{fig:comp_criptos}
    \end{subfigure}
    \hfill
    \begin{subfigure}{0.48\textwidth}
        \centering
        \includegraphics[width=\linewidth]{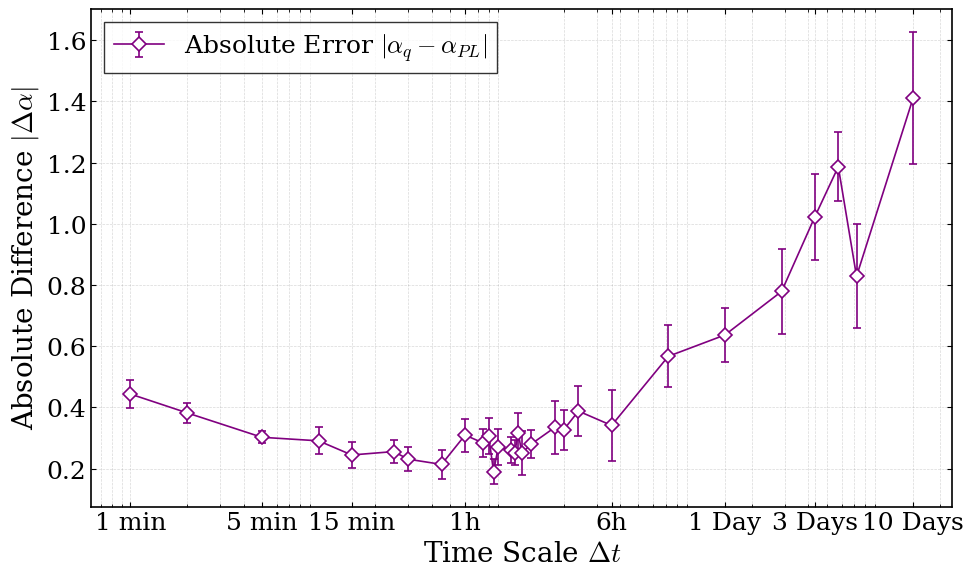}
        \caption{Absolute difference $|\alpha_q - \alpha_{PL}|$ for cryptocurrencies.}
        \label{fig:dif_criptos}
    \end{subfigure}
    
    \caption{Comparison of the evolution of the tail parameter $\alpha$ derived from pure power-law fits ($\alpha_{PL}$) and the asymptotic q-Gaussian approximation $\alpha = \frac{2}{q-1}$ ($\alpha_{q}$) across different aggregation scales. The individual evolution of each parameter and the absolute difference between them are presented for U.S. and Brazilian stock market, and cryptocurrencies.}
    \label{fig:grid_completo}
\end{figure}

Comparing the tail parameters derived from the pure power-law fit and the asymptotic q-Gaussian approximation reveals significant differences, as shown in Fig. \ref{fig:grid_completo}. A consistent pattern emerges across all three scenarios: at high-frequency scales, the q-Gaussian model underestimates $\alpha$ relative to the power law, predominantly in the region where the power law is shown to provide a better fit. At mesoscales (approximately 1 hour), both fits converge to similar values, reaching their lowest absolute difference.

At longer, interday scales, the q-Gaussian model captures aggregational Gaussianity, causing $q$ to approach 1 and $\alpha$ to increase, as the exponential decay characteristic of a Gaussian distribution would imply that the power-law exponent diverges to infinity ($\alpha \rightarrow \infty $). In contrast, the pure power-law fit fails to capture this effect, which leads it to underestimate the exponent $\alpha$ at these macroscopic scales. This demonstrates that even if both distributions are theoretically equivalent in the tail, the distinct fitting methodologies ultimately yield different exponents.

Finally, we believe that our contributions revisit and substantially extend some of the early investigations reported in \cite{tsallis1999,ramos1999intermittency}, which at that time were restricted to price changes in the U.S. dollar--German mark exchange rate. Here, the analysis is extended to stock markets and cryptocurrencies and is supported by a much finer and more detailed statistical investigation. These results may provide a useful framework for market analysts, scientists, investors, and other professionals working with economic and financial data. In particular, generalized functions based on ($q$) allow for a global description of the data, avoiding an a priori partition into distinct regions and the consequent need to identify distributional cutoffs before carrying out the analysis.

\vspace{0.3cm}

\textbf{Funding:} R. da Silva was partially supported by CNPq under grants 304575/2022-4 (Productivity Fellowship – PQ), 406820/2025-2 (Universal), 408596/2024-4 (INCT–NanoQuantSS), and 444867/2024-4 (CNPq/MCTI/FNDCT). J.R.Iglesias thanks to CNPq by the partial finantial support: 406820/2025-2 (Universal). R. da Silva also acknowledges financial support from FAPERGS (Fundação de Amparo à Pesquisa do Estado do Rio Grande do Sul) under Grant 25/2551-0002529-0.

\bibliographystyle{unsrt}
\bibliography{referenciasnew}

\end{document}